\documentclass[aps,twocolumn,prb,showpacs,preprintnumbers,superscriptaddress,amsmath,amssymb]{revtex4}

    \usepackage[dvips]{graphicx} 

\usepackage[english]{babel}
\usepackage{blindtext}
\usepackage{dcolumn}
\usepackage{bm}
\usepackage{ulem}
\usepackage[dvips]{color} 

\newcommand{\co}[2]{\ifcase #1 \or #2 \fi}

\newcommand{\bscco}{Bi$_{2}$Sr$_{2}$CaCu$_{2}$O$_{8}$\,}

\newif\ifnote

\begin{document}

\title{Thermal and electromagnetic properties of Bi$_2$Sr$_2$CaCu$_2$O$_8$ intrinsic Josephson junction stacks studied via one-dimensional coupled sine-Gordon equations}

\author{F.~Rudau}
\affiliation{Physikalisches Institut and Center for Collective Quantum Phenomena in LISA$^+$,Universit\"{a}t T\"{u}bingen, D-72076 T\"{u}bingen, Germany}
\author{M.~Tsujimoto}
\affiliation{Physikalisches Institut and Center for Collective Quantum Phenomena in LISA$^+$,Universit\"{a}t T\"{u}bingen, D-72076 T\"{u}bingen, Germany}
\affiliation{Kyoto University, Kyoto, Japan}
\author{B.~Gross}
\affiliation{Physikalisches Institut and Center for Collective Quantum Phenomena in LISA$^+$,Universit\"{a}t T\"{u}bingen, D-72076 T\"{u}bingen, Germany}
\author{T.E.~Judd}
\affiliation{Physikalisches Institut and Center for Collective Quantum Phenomena in LISA$^+$,Universit\"{a}t T\"{u}bingen, D-72076 T\"{u}bingen, Germany}
\author{R.~Wieland}
\affiliation{Physikalisches Institut and Center for Collective Quantum Phenomena in LISA$^+$,Universit\"{a}t T\"{u}bingen, D-72076 T\"{u}bingen, Germany}
\author{E.~Goldobin}
\affiliation{Physikalisches Institut and Center for Collective Quantum Phenomena in LISA$^+$,Universit\"{a}t T\"{u}bingen, D-72076 T\"{u}bingen, Germany}
\author{N.~Kinev}
\affiliation{Kotel'nikov Institute of Radio Engineering and Electronics, Moscow, Russia}
\author{J.~Yuan}
\affiliation{National Institute for Materials Science, Tsukuba 3050047, Japan}
\author{Y.~Huang}
\affiliation{National Institute for Materials Science, Tsukuba 3050047, Japan}
\affiliation{Research Institute of Superconductor Electronics, Nanjing University, Nanjing 210093, China}
\author{M.~Ji}
\affiliation{National Institute for Materials Science, Tsukuba 3050047, Japan}
\affiliation{Research Institute of Superconductor Electronics, Nanjing University, Nanjing 210093, China}
\author{X.J.~Zhou}
\affiliation{National Institute for Materials Science, Tsukuba 3050047, Japan}
\affiliation{Research Institute of Superconductor Electronics, Nanjing University, Nanjing 210093, China}
\author{D.Y.~An}
\affiliation{National Institute for Materials Science, Tsukuba 3050047, Japan}
\affiliation{Research Institute of Superconductor Electronics, Nanjing University, Nanjing 210093, China}
\author{A.~Ishii}
\affiliation{National Institute for Materials Science, Tsukuba 3050047, Japan}
\author{R.G.~Mints}
\affiliation{The Raymond and Beverly Sackler School of Physics and Astronomy, Tel
Aviv University, Tel Aviv 69978, Israel}
\author{P.H.~Wu}
\affiliation{Research Institute of Superconductor Electronics, Nanjing University, Nanjing 210093, China}
\author{T.~Hatano}
\affiliation{National Institute for Materials Science, Tsukuba 3050047, Japan}
\author{H.B.~Wang}
\affiliation{National Institute for Materials Science, Tsukuba 3050047, Japan}
\author{V.P.~Koshelets}
\affiliation{Kotel'nikov Institute of Radio Engineering and Electronics, Moscow, Russia}
\author{D.~Koelle}
\affiliation{Physikalisches Institut and Center for Collective Quantum Phenomena in LISA$^+$,Universit\"{a}t T\"{u}bingen, D-72076 T\"{u}bingen, Germany}
\author{R.~Kleiner}
\affiliation{Physikalisches Institut and Center for Collective Quantum Phenomena in LISA$^+$,Universit\"{a}t T\"{u}bingen, D-72076 T\"{u}bingen, Germany}
\date{\today}

\begin{abstract}
We used  one-dimensional coupled sine-Gordon equations combined with heat diffusion equations to numerically investigate the thermal and electromagnetic properties of a 300\,$\mu$m long intrinsic Josephson junction stack consisting of $N$ = 700 junctions. The junctions in the stack are combined to $M$ segments where we assume that inside a segment all junctions behave identically. Most simulations are for $M = 20$. For not too high bath temperatures there is the appearence of a hot spot at high bias currents. In terms of electromagnetic properties, robust standing wave patterns appear in the current density and electric field distributions. These patterns come together with vortex/antivortex lines across the stack that correspond to $\pi$ kink states, discussed before in the literature for a homogeneous temperature distribution in the stack. 
We also discuss scaling of the thermal and electromagnetic properties with $M$, on the basis of simulations with $M$ between~10~and~350.
\end{abstract}



\maketitle

\section{Introduction}
\label{sec:intro}
In 2007 it has been shown\cite{Ozyuzer07} that stacks of intrinsic Josephson junctions (IJJs)\cite{Kleiner92} in the high temperature superconductor \bscco (BSCCO) are sources of coherent radiation at THz frequencies, with the possibility to tune the emitted frequency $f_{\rm e}$ by an applied dc voltage $V$, following the relation $f_{\rm e} = V/\Phi_0$. Here $\Phi_0$ is the flux quantum and $\Phi^{-1}_0$ = 483.6\,GHz/mV. In Ref. \onlinecite{Ozyuzer07} stacks of about 1\,$\mu$m in thickness (corresponding to 666 IJJs), a length $L_{\rm s}$ of about 300\,$\mu$m and a width $W$ of some 10\,$\mu$m have been realized as mesa structures on top of BSCCO single crystals, contacted by Au layers. These mesas emitted radiation at frequencies between 0.5 and 0.8\,THz, with an integrated output power on the order of 1\,$\mu$W. The emission frequency was found to scale reciprocally with $W$, indicating that cavity modes, formed along the width of the stack,  are responsible for synchronization. 

THz radiation emitted from such IJJ stacks became a hot topic in recent years, 
both in terms of experiment 
\cite{Wang09a, Minami09, Kurter09,Gray09,Guenon10, Kurter10, Wang10a, Tsujimoto10,Koseoglu11,Benseman11,Yamaki11,Yuan12,Li12,Wang12,Tsujimoto12,Kakeya12,Tsujimoto12a,Turkoglu12,Oikawa13, An13,Benseman13,Benseman13a,Benseman13b,Minami14,Watanabe14,Ji14}
and theory
\cite{Bulaevskii07, Koshelev08,Koshelev08b,Lin08,Krasnov09,Klemm09,Nonomura09,Tachiki09,Pedersen09,Hu09,Koyama09,Grib09,Zhou10,Krasnov10,Koshelev10,Savelev10,Lin10a,Lin10b,Lin10c,Katterwe10,Yurgens11,Koyama11,TachikiT11,Slipchenko11,Krasnov11,Yurgens11b,Lin11b,Asai12,Asai12b,Zhang12,Lin12,Averkov12,Grib12,Gross12,Apostolov12,Liu13,Asai13,Asai14,Grib14,Lin14}. For recent reviews, see Refs. \onlinecite{Kashiwagi12,Welp13, Kawayama13}.
IJJ stacks, containing typically 500 -- 2000 junctions, have been realized as mesa structures but also as bare IJJ stacks contacted by Au layers \cite{Kashiwagi12,An13,Sekimoto13,Ji14} and as all-superconducting z-shaped structures \cite{Yuan12}. Emission frequencies are in the range 0.4 -- 1\,THz. For the best stacks, emission powers in the range of tens of $\mu$W have been achieved \cite{An13,Sekimoto13,Benseman13}, and arrays of stacks showed emission with a power up to 0.61\,mW \cite{Benseman13a}.  

A crucial point in the physics of the huge IJJ stacks is overheating \cite{Kurter09,Kurter10,Wang09a,Wang10a,Guenon10,Kakeya12,Yurgens11, Yurgens11b,Asai12,Gross12,Benseman13,Minami14,Grib14,Asai14}. For sufficiently low bias currents, the temperature rises only slightly to values above the bath temperature $T_{\rm{bath}}$ and the voltage across the stack $V$ increases with increasing bias current $I$. With increasing $I$ and input power the current voltage characteristics (IVCs) start to back-bend and, at some bias current in the back-bending region, a hot spot forms suddenly in the stack \cite{Wang09a,Wang10a,Guenon10,Kakeya12,Benseman13,Minami14,Watanabe14}, creating a region which is heated to temperatures above the critical temperature $T_{\rm c}$. The reason is the strong increase of the BSCCO $c$-axis resistivity $\rho_{c}$ with decreasing temperature together with the poor BSCCO thermal conductivity \cite{Yurgens11, Gross12, Gross13}. Similar effects also occur in other systems \cite{Gurevich87, Spenke36b}.  In the IJJ stacks one can thus distinguish a low-bias regime where the temperature in the mesa varies only weakly and a high-bias regime where the hot spot has formed, leaving the ``cold'' part of the mesa for THz generation via the Josephson effect. The formation of the hot spot also affects the THz emission properties of the stack. For example, it has been found that the linewidth of radiation is much more narrow in the high-bias regime than at low bias \cite{Li12}. This can be reproduced by simple model calculations based on arrays of pointlike junctions \cite{Gross13}.  On the other hand several other properties like the emission frequency seem to be basically independent on the hot spot position. This has lead to some debate whether the hot spot is helpful or just coexists with the electromagnetic properties \cite{Sekimoto13,Minami14,Watanabe14}.

In terms of theory many calculations of electrodynamics have been based on a homogeneous temperature distribution within a stack, while calculations of the thermal properties were based on solving the heat diffusion equations in the absence of Josephson currents \cite{Yurgens11,Yurgens11b,Gross12}. Some attempts have been made to combine both electrodynamics and thermodynamics, either by using arrays of pointlike IJJs \cite{Gross13,Grib14} or by incorporating temperature induced effects into an effective model describing the whole stack as a single ``giant'' junction \cite{Asai12,Asai13,Asai14}. As we will see the latter approach has inconsistencies. 

In this paper we report on simulations where we solve the one-dimensional coupled sine-Gordon equations in combination with the heat diffusion equations. In our approach we group the junctions in the stack to segments. We still assume that all junctions in a segment behave like a giant junction. We find many thermodynamic and electromagnetic properties that have been seen in experiment and also in the previous theoretical calculations, but also there are new features. Despite the good agreement with several experimental observations we cannot emphasize strongly enough that our approach is still far from the 3D case where all junctions in the stack are addressed individually and where in-plane variations of the thermal and electromagnetic properties are taken into account in 2D.

The reminder of the paper is organized as follows. In Sec. \ref{sec:model} and in the appendix we introduce the geometry considered, together with the basic equations, the simplifications made and the numerical procedures used. In Sec. \ref{sec:results} we present our results, starting with integral properties like IVCs, then turning to local properties and finally commenting on scaling issues and some special properties like the role of the hot spot position and the observability of low temperature scanning laser microscopy (LTSLM) signals. We conclude and summarize in Sec. \ref{Sec:Conclusions}.


\section{Model}
\label{sec:model}

\subsection{Geometry and basic equations}

\begin{figure*}[tb]
\includegraphics[width=0.8\textwidth,clip]{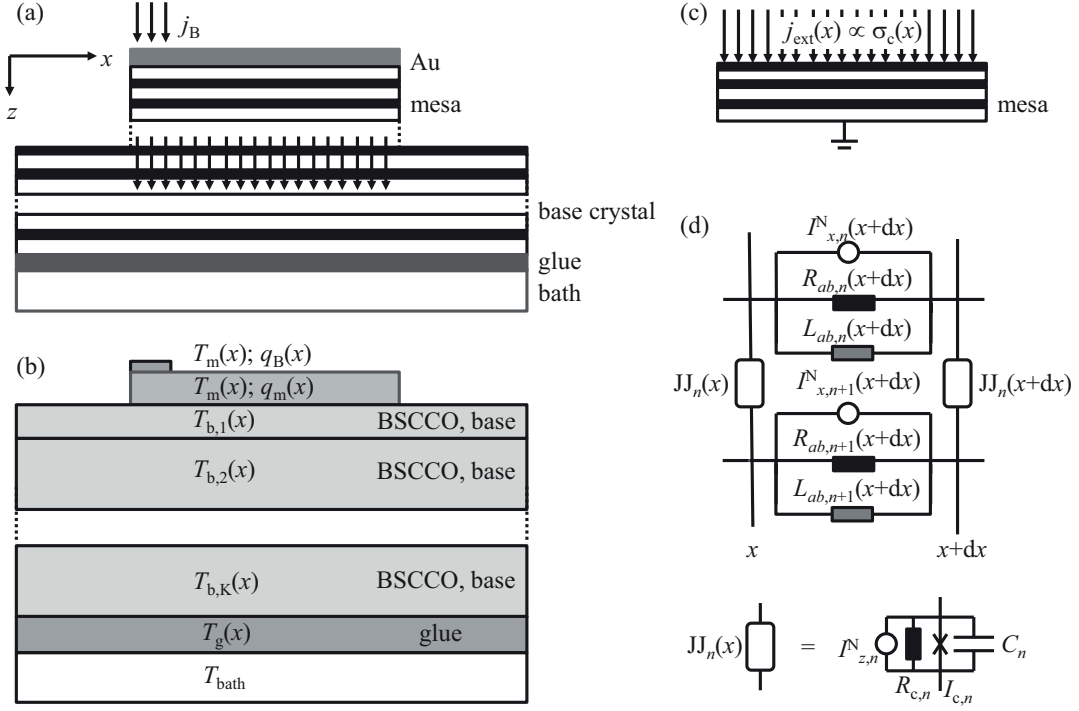}
\caption{General geometry and simplifications used for modelling. (a) Sketch of mesa geometry. (b) Sketch of geometry for thermal description.  The Joule power density $q_{\rm m}$ is produced in the mesa and $q_{\rm B}$ is the Joule power density produced by the bias lead. 
The temperatures of the various layers are indicated. (c) Simplified geometry considered for the electrical description together with (d) the lumped circuit approximation. In (d) one element describing the mesa between positions $x$ and $x+$d$x$ and the IJJ $n$ embedded between the superconducting layers $n$ and $n+1$ is shown. The in-plane currents in the $n$th layer are approximated by an inductor $L_{{ab},n}$ (supercurrent), a resistor $R_{{ab},n}$ (quasiparticle current) and a noise source $I^{\rm N}_{{x},n}$. 
The interlayer current is described by the Josephson current with critical current $I_{{\rm c},n}$, a resistor $R_{{c},n}$, a capacitor $C_n$ and a noise source $I^{\rm N}_{{z},n}$
}
\label{fig:geometry}
\end{figure*}

We consider an IJJ stack (mesa) consisting of $N$ IJJs, cf. Fig.~\ref{fig:geometry}(a). The thickness of the superconducting layers (CuO$_2$ planes) is\cite{Kleiner92} $d_{\rm s}$ = 0.3\,nm and the thickness of the insulating layers between the CuO$_2$ planes is $d_{\rm i}$ = 1.2\,nm. The stack has a length $L_{\rm s}$ along $x$ and a width $W$ along $y$. The mesa thickness is $D_{\rm m} = Ns$, where $s = d_{\rm s} + d_{\rm i}$. All electrical and thermal properties shall be homogeneous along $y$. The mesa is covered by a gold layer of thickness $D_{\rm{Au}}$ and is patterned on a base crystal of length $L_{\rm b} > L_{\rm s}$ and thickness $D_{\rm b}$. The base crystal is mounted by a glue layer of thickness $D_{\rm g}$ to a sample holder which is kept at a bath temperature $T_{\rm{bath}}$. A bias current $I$ is injected via a bond wire into the Au layer and leaves the mesa into the base crystal. Note that the electrical and thermal parameters (resistivities, critical current densities, thermal conductivities etc.) introduced below depend on temperature, and thus, for an inhomogeneous temperature distribution, on $x$. Their spatial variation, as well as $T(x)$, can be found by self-consistently solving the thermal equations (requiring Joule heat dissipation as an input from the electric circuit) and the electrical equations (requiring $T(x)$, as determined from the thermal circuit, as an input).

For the thermal description (cf. Fig.~\ref{fig:geometry}(b)) we assume that the mesa plus the contacting Au layer and the bond wire have a temperature $T_{\rm m}(x)$ which is constant along $z$ but can vary along $x$. The effective thickness of this layer is $D_{\rm{m,eff}}$. 
The BSCCO base crystal is split into $K$ segments. The segment interfacing the mesa also has the thickness $D_{\rm{m,eff}}$, the other layers have a (much larger) thickness $(D_{\rm b}-D_{\rm{m,eff}})/(K-1)$. The temperature in the center (along $z$) of the $k$th segment of the base crystal is $T_{{\rm b},k}(x)$, and the temperature in the center of the glue layer is $T_{\rm g}(x)$. The whole ensemble is coupled in $z$ direction to the bath which is defined to have a constant temperature $T_{\rm{bath}}$.

Generally speaking, we solve the heat flow equation
\begin{equation}
	\label{eq:heat_diffusion}
	c\dot{T}=\nabla\left(\kappa\nabla T\right)+q
\end{equation}
with the specific heat $c$, the (anisotropic) thermal conductivity $\kappa$ and the power density for heat generation $q$. The dot denotes derivative with respect to time. Eq.~(\ref{eq:heat_diffusion}) needs to be specified for the layered geometry of Fig.~\ref{fig:geometry}(b).
Details are given in the appendix. In brief, we solve in the $k$th layer ($k$ runs from 0 to $K$+1 and includes the mesa ($k$ = 0) and the glue layer ($k$ = $K$ + 1)):
\begin{equation}
\label{eq:heat_nth_layer}
c_k\dot{T}_k=\frac{d}{dx}\left(\kappa_{\|,k}\frac{d}{dx}T_k\right)+\frac{2}{D_k}(j_{{\rm in},k}-j_{{\rm out},k})+q_k,
\end{equation}
where $j_{{\rm{in}},k}$ and $j_{{\rm out},k}$, respectively, denote the heat current densities into and out of the layer $k$. $\kappa_{\|,k}$ is the in-plane thermal conductivity of layer $k$, $T_k$ is its temperature and $D_k$ is its thickness; $c_k$ is the heat capacity of layer $k$. For layer 0 (mesa plus gold plus bond wire) $q_0$ denotes the Joule power density $q_{\rm m}$ produced by the in-plane and out-of-plane currents in the mesa, plus the power density $q_{\rm B}$ produced by the bond wire. The latter contribution has turned out to be very useful in the simulations since, for high enough $q_{\rm B}$, the hot spot forming in the mesa is located near the wire position. In the layers representing the base crystal and the glue there is no heat generation, i.e. $q_k = 0$ here. These layers have a length $L_{\rm b}$ which we have taken as 2$L_{\rm s}$. The mesa is centered above the base crystal.  

The electric circuit is sketched in Fig.~\ref{fig:geometry}(c). We have grouped the $N$ IJJs in the stack to $M$ segments, each containing $G = N/M$ IJJs, assumed to have identical properties. 

The mesa is biased by an external current density $j_{\rm{ext}}(x)$ which enters the mesa in $z$ direction with a density proportional to the local BSCCO conductance $\sigma_{c} (x) = \rho_c^{-1}(x)$, i.e. we assumed that the Au layer has a low enough resistance to freely distribute the current injected by the bond wire along $x$ before it enters the IJJ stack.  The interface of the stack to the base crystal is treated as a ground. 

Fig.~\ref{fig:geometry}(d) shows the lumped circuit approximaton for a piece of the \textit{single} IJJ $n$, located between $x$ and $x$ + d$x$.  
For the current flow along $z$, we consider a Josephson current with critical current $I_{{\rm c},n}$, a resistive component with $R_{{c},n}$ and a capacitive component $C_n$. Nyquist noise is considered via  a random current source $I^{\rm N}_{{z},n}$ with spectral power density $4k_{\rm B}T_{\rm m}/R_{{c},n}$.
The in-plane current flow in the $n$th BSCCO layer is described by a resistive component $R_{{ab},n}$  and an inductive component $L_{{ab},n}$ which  is the kinetic inductance associated with in-plane supercurrents. We also consider an in-plane noise current 
$I^{\rm N}_{{x},n}$ with a spectral power distribution  $4k_{\rm B}T_{\rm m}/R_{{ab},n}$. 

As described in the appendix this leads to a sine-Gordon-like equation for the $m$th segment of the IJJ stack:

\begin{equation}
\label{eq:sigo_segment}
\begin{split}
Gsd_{\rm s}\left(\frac{\dot{\gamma}^\prime_m}{\rho_{ab}}\right)^\prime +d_{\rm s}\left(j^{\rm N}_{{x},m+1}-j^{\rm N}_{{x},m}\right)^\prime + G\lambda_{\rm k}^2 \left(n_s\gamma^\prime_m\right)^\prime = \\
2j_{z,m}-j_{z,m+1}-j_{z,m-1}.
\end{split}
\end{equation}
The index $m$ runs from 1 to $M$ and enumerates the $M$ segments. 
The characteristic length $\lambda_{\rm k} = [\Phi_0 d_{\rm s}/(2\pi\mu_0j_{\rm c0}\lambda_{ab0}^2)]^{1/2}$, with the 4.2\,K value of the in-plane London penetration depth $\lambda_{ab0}$ and the magnetic permeability $\mu_0$. The in-plane resistivity is denoted $\rho_{ab}$ and $n_{\rm s} = \lambda^2_{ab0}/\lambda^2_{ab}$ denotes the Cooper pair density.  
Time is normalized to $\Phi_0/2\pi j_{\rm c0}\rho_{c0}s$, resistivities to $\rho_{c0}$ (4.2\,K value of $c$-axis resistivity) and current densities to $j_{\rm c0}$ (4.2\,K value of Josephson current critical density). The primes denote derivative with respect to $x$ and $\gamma_m$ is the Josephson phase difference of each IJJ in segment $m$. We have further assumed that resistivites and critical current densities are the same for all layers, i.e. do not depend on $m$. For the in-plane noise current the normalized form of the spectral density is  $4\Gamma_0 (T_{\rm m}/T_0)d_{\rm s}s/($d$xL_{\rm s}\rho_{ab}$), with $T_0 = 4.2\,K$, $\Gamma_0 = 2\pi k_{\rm B}T_0/I_{\rm c0}\Phi_0$ and $I_{\rm c0} = j_{\rm c0}WL_{\rm s}$.

For the current densities $j_{{z},m}$ one finds
\begin{equation}
\label{eq:RCSJ}
j_{{z},m} = \beta_{\rm c0} \ddot{\gamma}_m + \frac{\dot{\gamma}_m}{\rho_{{c},m}} + j_{\rm c} \sin(\gamma_m) +j^{\rm N}_{{z},m},
\end{equation}
with $\beta_{\rm c0} = 2\pi j_{\rm c0}\rho_{c0}^2\epsilon\epsilon_0s/\Phi_0$; $\epsilon_0$ is the vacuum permittivity and $\epsilon$ is the BSCCO dielectric constant.
The normalized spectral density of $j^{\rm N}_{{z},m}$ is $4\Gamma_0 (T_{\rm m}/T_0)L_{\rm s}/($d$x\rho_{c}$).  
 
The in-plane supercurrent densitities in electrode $m$ (the CuO$_2$ layer interfacing segments $m$ and $m+1$) are expressed as
\begin{equation}
\label{eq:jxs}
j^{\rm s}_{{x},m} = \frac{\lambda_{\rm k}^2}{d_{\rm s}}n_{\rm s}\phi_m^\prime
\end{equation}
where $\phi_m$ is the phase of the superconducting wave function in electrode $m$.
The resistive currents in electrode $m$ are given by
\begin{equation}
\label{eq:jxr}
j^{\rm r}_{{x},m} = \frac{s}{\rho_{ab} } \dot{\phi}_m^\prime
\end{equation}
and the $\phi_m$ and $\gamma_m$ are related via
\begin{equation}
\label{eq:phi_gamma}
\gamma_m = \frac{\phi^\prime_{m+1}-\phi^\prime_{m}}{G}, 
\end{equation}
allowing to evaluate the in-plane currents once all $\gamma_m$ and in addition $\phi_m^\prime$ of one of the outermost electrodes are known.   

The expression for the current density $j_{\rm{ext}}$ is
\begin{equation}
\label{eq:j_ext}
j_{\rm{ext}} = \frac{\left\langle j_{\rm{ext}}\right\rangle}{\left\langle\sigma_{c}\right\rangle\rho_{c}},
\end{equation}
where the brackets denote spatial averaging.

Note that Eqs. (3) and (4) together with Eq. (2) have the same form as the equations for a stack of $M$ single junctions rather than $M$ segments. On the thermal side the difference is that the $M$ segments produce the same Joule heat as the full $N$ junction stack (a stack of, say, 20 junctions would not heat up significantly). On the electromagnetic side the first difference is the rescaling of the length $\lambda_k$ which is multiplied by $G^{0.5}$, as well as a rescaling of the in-plane resistance $\rho_{ab}$ which is divided by $G$ (see first and third term on the left hand side of Eq. 2). As we will see in section III there are robust in-phase standing waves along the stack. The mode velocity of these waves increases $M$ and, in order to keep physics independent of the segmentation we kept the product $\beta_{c0} G$ constant. This modification is discussed and justified in detail in the appendix.  
With these scalings the electrothermal properties calculated are very similar for all values of $G$ and follow simple scaling rules, as shown in section III.

%
\begin{figure}[tb]
\includegraphics[width=\columnwidth,clip]{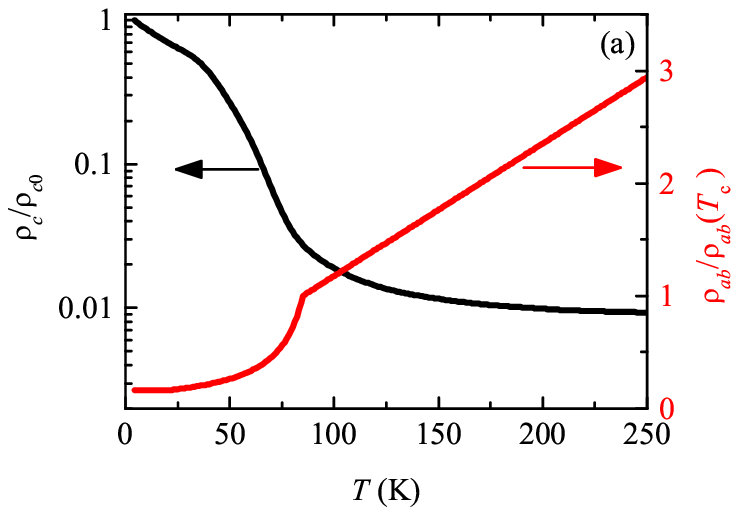}\\
\includegraphics[width=\columnwidth,clip]{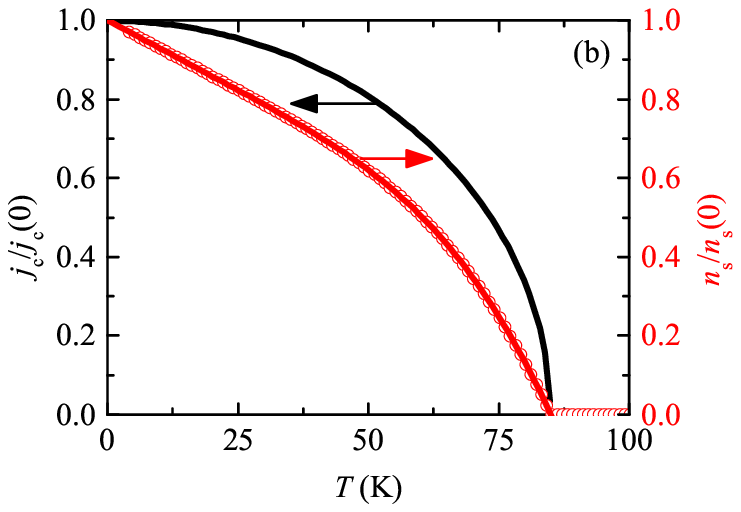}
\caption{(color online) Temperature dependences of various parameters used for calculations. (a) out-of-plane (left scale) and in-plane (right scale) resistivities. (b) Josephson critical current density (left scale) and superfluid density (right scale). 
}
\label{fig:parameters}
\end{figure}
%

\subsection{Choice of parameters} 
We perform our calculations for a $N$ = 700 IJJ mesa with lateral dimension $L_{\rm s}$ = 300\,$\mu$m. The length of the base crystal is $L_{\rm b}$ = 600\,$\mu$m and its thickness is $D_{\rm b}$ = 30\,$\mu$m. The mesa is centered above the base crystal. The thickness of the glue layer is $D_{\rm g}$ = 20\,$\mu$m. The BSCCO critical temperature is $T_{\rm c} = 85$\,K.

For $\rho_{c} (T)$ we take a 4.2\,K value of 10$^3$\,$\Omega$cm and the temperature dependence used in Ref. \onlinecite{Gross12}. Above the transition temperature $T_{\rm c}$ it is based on measured data. Below $T_{\rm c}$, $\rho_{c}$ is extrapolated to give good agreement to measured IVCs. Fig.~\ref{fig:parameters}(a) shows this functional form.
We have based the BSCCO in-plane resistivity $\rho_{ab}$ on microwave surface impedance measurements \cite{Lee96}. In general, the real part of the in-plane conductivity $\sigma_1$ below $T_{\rm c}$ runs over a low-temperature maximum and in addition is likely to be strongly frequency dependent \cite{Nunner05}. Being mostly interested in temperatures well above 20\,K we linearized this quantity for temperatures between 20\,K and $T_{\rm c}$ and took $\sigma_1$ as a constant below 20\,K. For temperatures above $T_{\rm c}$ we assumed that $\rho_{ab}$ increases linearly with temperature. 
This yields $\rho_{ab} (T)/\rho_{ab}(T_{\rm c}) = T/T_{\rm c}$ for temperatures above $T_{\rm c}$, and $\rho_{ab} (T)/\rho_{ab}(T_{\rm c}) = (1+a(T_{\rm c}-T))^{-1}$ for $20\,K < T < T_{\rm c}$ 
and $\rho_{ab} (T)/\rho_{ab}(T_{\rm c}) = (1+a(T_{\rm c}-20$\,K$))^{-1}$ for $T < 20$\,K. For $a$ we used a value of 0.08\,K$^{-1}$. The resulting curve is shown in Fig.~\ref{fig:parameters}(a).
We further used $\rho_{ab} (T_{\rm c})$ = 20\,$\mu\Omega$cm. This value sounds somewhat low. However, we assume that only layers of thickness $d_{\rm s} = 0.3$\,nm are conducting while the interlayers of thickness $d_{\rm i} = 1.2$\,nm are insulating. This results in an averaged in-plane resistivity of 100\,$\mu\Omega$cm, which is realistic. 
 
For the Josephson critical current density at $T$ = 4.2\,K we use $j_{\rm c0} = 200$\,A/cm$^2$. For the temperature dependence of $j_{\rm c}$, for $T < T_{\rm c}$ = 85\,K we use (cf. Fig.~\ref{fig:parameters}(b))
\begin{equation}
\label{eq_A:jcT}
j_{\rm c}(T)= j_{\rm c}(0) \left[1-(T/T_{\rm c})^2\right]^{1/2}.
\end{equation}
For $T > T_{\rm c}$, $I_{\rm c}(T)$ = 0. Eq.~(\ref{eq_A:jcT}) roughly approximates $I_{\rm c}(T)$ data of small-sized BSCCO stacks\cite{Kleiner92}. 

The superfluid density $n_{\rm s}(T) \propto \lambda^2_{ab}(T)/\lambda^2_{ab}(0)$ is taken from Ref. \onlinecite{Jacobs95}. The temperature dependence of this curve can be fitted very well by  
\begin{equation}
\label{eq_A:nsT}
n_{\rm s}(T)= n_{\rm s}(0) \big[1-(T/T_{\rm c})^6\big]\big[1-0.6(T/T_{\rm c})\big].
\end{equation}
Fig.~\ref{fig:parameters}(b) shows by points data from Ref. \onlinecite{Jacobs95} and by a line the fit function, Eq.~(\ref{eq_A:nsT}). For $\lambda_{ab} (0)$ we have used a value of 260\,nm.

With the above choice of parameters, assuming a mesa width of $W$ = 50\,$\mu$m, we obtain the following 4.2\,K values for the electrical part of the model. Critical current $I_{\rm c0}$ = 30\,mA, $c$-axis resistance per junction $R_{c0}$ = 1\,$\Omega$, $I_{\rm c0}R_{c0}$ = 30\,mV, characteristic frequency $f_{\rm c0}$ = 14.5\,THz and  noise parameter $\Gamma_0 \approx$ 5$\cdot10^{-6}$.
The characteristic power density $p_{\rm c0} = j^2_{\rm c0}\rho_{c0}$ is $4\cdot10^7$\,W/cm$^3$, yielding, for a stack volume of $1.5\cdot10^{-8}$\,cm$^3$, a characteristic power $P_{\rm c0}$ of 0.6\,W. For $\lambda_{c}$ one obtains 296\,$\mu$m and  $\lambda_{\rm k} = 0.76$\,$\mu$m. 

Further, assuming a (temperature independent) dielectric constant $\epsilon = 12$ (diffraction index 3.5) we obtain for the Josephson plasma frequency $f_{\rm pl0} \approx$ 41\,GHz. The McCumber parameter (for $G$ = 1) is $\beta_{\rm c0} \approx 1.25\cdot10^5$. The 4.2\,K value of the in-phase mode velocity $c_1$ (see Eq.~(\ref{eq_A:mode_velocity})) is $8.8\cdot10^7$\,m/s. $c_1$ decreases $\propto n_{\rm s}^{1/2}$ with temperature.  
In our simulations we keep the product $\beta_{\rm c0}G$ constant in order to (approximately) fix the 4.2\,K value of $c_1$.
Still, $c_1$ slightly increases with increasing $G$. To compensate for this we have used $\beta_{\rm c0}G = 1.4\cdot10^5$ for calculations with $G$ between 14 and 70,  cf. appendix, Fig.~\ref{fig_A:mode_velocities}.

For the BSCCO thermal conductivities we use 4.2\,K values $\kappa_{ab}$ = 2.76\,W/mK and $\kappa_{c}$ = 0.32\,W/mK.
For Au we use $\kappa_{\rm Au}$ = 100\,W/mK 
and for the glue $\kappa_{\rm g}= $ 0.5\,W/mK; the two latter values were taken as temperature independent for simplicity.
The heat capacities, determining the time dependence of establishing temperature distributions in the various layers we simply kept constant, keeping in mind that we are, for the moment, mainly interested in the Josephson dynamics which is much faster than the dynamics of the thermal part. I.e., although solving dynamic equations, we are interested in situations where the temperature distributions are basically stationary. We thus used a heat capacity per volume of 2\,J/m$^3$K for all layers.

For the heat produced by the bond wire we used $\rho_{\rm B} = 0.02\,\rho_{c0}$ and a diameter (along $x$) $L_{\rm B} = 30$\,$\mu$m.

\subsection{Numerical details and quantities calculated}
Equations (\ref{eq:heat_nth_layer}) and (\ref{eq:sigo_segment}) were discretized along $x$ using equally spaced grid points. A 5th order Runge-Kutta scheme was used to evolve these equations in time.  We reduced the number of grid points $X$ along the stack as much as possible to speed up the calculations; $X$ = 50 was used for the simulations shown and some of the results were confirmed using $X$ = 100. For the base crystal and the glue layer 2$X$ grid points were used. The base crystal was split into $K$ = 4 segments.  
For a given set of input parameters, in a first initializing step we solved, for typically $10^9$ time units, Eq.~(\ref{eq:heat_nth_layer}) for out-of-plane quasiparticle currents only to achieve stationary distributions for the temperature and $j_{\rm{ext}}$. Then, in a second initializing step, Eq.~(\ref{eq:heat_nth_layer}) was solved simultaneously with Eq.~(\ref{eq:sigo_segment}) over typically $10^4/v$ time units. Here, $v$ is the temperature dependent normalized averaged dc voltage across a junction, $v = V/(NV_{\rm c0}$) (we divided the integration time by $v$ to keep the number of Josephson oscillations constant).  
After this second initializing step which is necessary to bring the electric circuit into a stationary state the $M\cdot X$ Josephson phases $\gamma_m(x)$ and also other quantities like $\dot{\gamma}_m(x)$, in-plane and out-of-plane current densities etc. were tracked as a function of time to produce time averages of these quantities or to make Fourier transforms. To obtain (hysteretic or even multibranched) current voltage characteristics (IVCs) one often starts at, say, zero bias current and then varies this current using the values for the various variables from the previous current for initialization. We did not use this concept but instead initialized each current, as described above, to obtain reproducible states in different runs. 

Apart from calculating IVCs and distributions of the thermal and electrical quantities in the mesa we were also interested in THz emission. 
In experiment one finds that a significant power is emitted in $z$ direction\cite{Kadowaki10,Kashiwagi12},  which must be due to currents oscillating in the CuO$_2$ planes. In our simulations we do not calculate electric and magnetic fields outside the IJJ stack and thus do not have access to the Poynting vector to calculate emission properties\cite{Koyama11}. Instead, we start from the power density $q_{x,{\rm av}}$ produced by the resistive part of the in-plane currents $j^{\rm r}_{x,m}$, averaged over the stack volume.
It is given by 
\begin{equation}
\label{eq:x_power}
q_{x,{\rm av}} = \frac{d_{\rm s}}{Ns}\cdot\frac{1}{L_{\rm s}}\int_0^{L_{\rm s}}dx \Big(\rho_{ab}\sum_{m=1}^{M+1}j^{\rm r 2}_{x,m}\Big)
\end{equation}
Because of our boundary conditions $j^{{\rm r} 2}_{x,m}$ and thus also $q_{x,{\rm av}}$ have no dc component. Next, we take time traces of $q_{x,{\rm av}}$, Fourier transform them and look at the peak in $q_{x,{\rm av}}(f)$ which occurs at twice the Josephson frequency. We denote the peak value as $q_{x{\rm p}}$.
Our normalization power densities are in units of $j^2_{\rm c0}\rho_{c0}$ which, integrated over the stack volume yields a characteristic power of 0.6\,W. Thus, $q_{x{\rm p}}$ can also be viewed as the in-plane power in the stack in units of 0.6\,W.  Of course, $q_{x{\rm p}}$ is the \textit{dissipated} rather than the \textit{emitted} power. However, it seems natural that both quantities track each other.

\section{Results}
\label{sec:results}
\subsection{Current voltage characteristics and in-plane power}
%
\begin{figure}[tb]
\includegraphics[width=\columnwidth,clip]{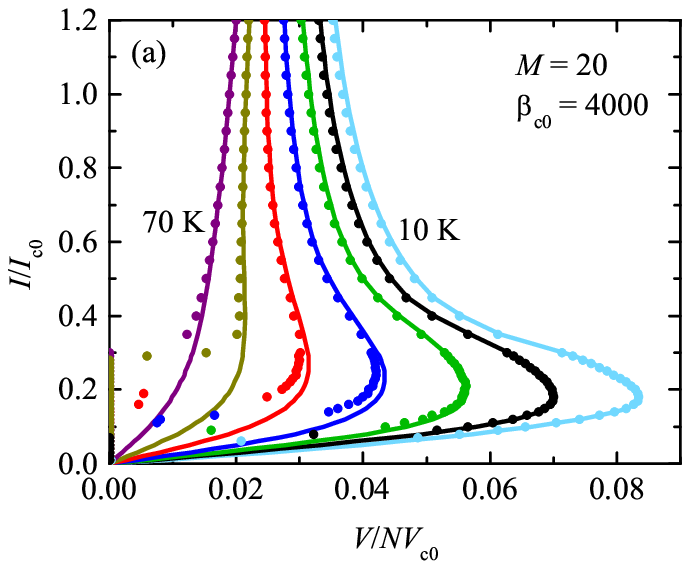}\\
\includegraphics[width=\columnwidth,clip]{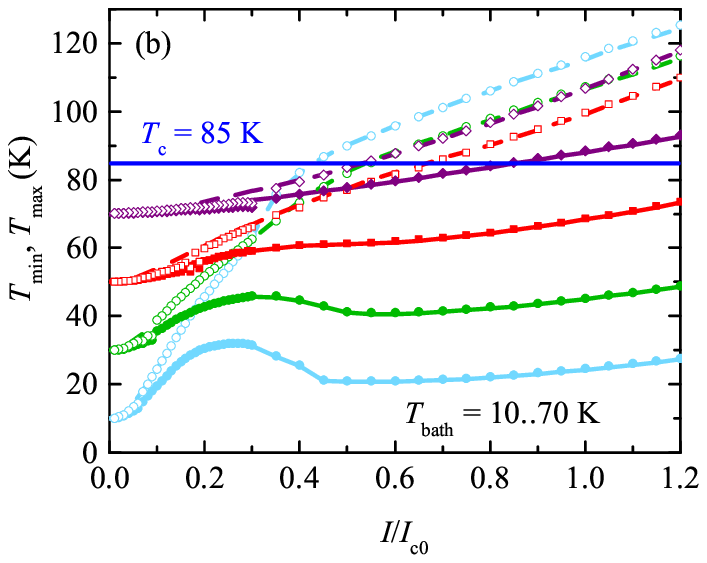}
\caption{(color online) Simulation results for an $M$ = 20 segment stack, with $G = N/M = 35$ and $\beta_{\rm c0} = 4000$ for bath temperatures between 10\,K and 70\,K. (a) IVCs after first initialization step, with only resistive $c$-axis currents taken into account (lines). Full calculation including Josephson currents, displacement currents and in-plane currents (dots). (b) Maximum and minimum temperatures in the stack vs. normalized bias current Lines (solid for $T_{\rm min}$, dashed for $T_{\rm max}$): after first initialization step; symbols (solid for $T_{\rm min}$, open for $T_{\rm max}$): full calculation. }
\label{fig:IVCs}
\end{figure}
%
We first discuss dc characteristics of an $M$ = 20 segment stack, with $G = N/M = 35$ and $\beta_{\rm c0} = 4000$ ($G\beta_{\rm c0} = 1.4\cdot 10^5$). Data are for bath temperatures between 10\,K and 70\,K.
The bias lead had a diameter along $x$ of $L_{\rm B}$ = 30\,$\mu$m and its left edge was positioned 30\,$\mu$m from the left edge of the mesa.
Fig.~\ref{fig:IVCs}(a) shows by solid lines IVCs, as they have been calculated in the first initialization step. Here, current flow is assumed to be purely in $z$ direction and only resistive currents (with resistance $\rho_{c}$) are taken into account in the heat diffusion equations. The IVCs show the typical back-bending which is due to self-heating. By points we indicate calculated IVCs using the full model equations (heat diffusion plus coupled sine-Gordon equations). For each bias current the $c$-axis electric fields in each segment ($\propto \dot{\gamma}_m$) have been initialized so that all IJJs are in their resistive state.
For not too low bias currents the points are basically on top of the lines showing that dissipation due to $c$-axis quasiparticle currents is the dominating effect for self-heating. At low bias currents there is a switch-back to either some ``inner'' branch of the IVC (some junctions resistive, some others in zero voltage state) or to the zero voltage branch. We note that, for a given IVC, inner branches can be traced out in principle, although we have not done this in the IVCs shown. Fig.~\ref{fig:IVCs}(b) displays the maximum ($T_{\rm{max}}$) and minimum ($T_{\rm{min}}$) temperatures in the stack as a function of bias current. Data are for bath temperatures between 10\,K and 70\,K, in steps of 20\,K. By lines we show $T_{\rm{max}}$ and $T_{\rm{min}}$ as calculated in the first initialization step. At low temperatures and currents both $T_{\rm{max}}$ and $T_{\rm{min}}$ increase with increasing current. With further current increase $T_{\rm{min}}$ runs over a maximum, while $T_{\rm{max}}$ exhibits a strong increase. This is a typical signature of hot spot formation starting at the maxima of $T_{\rm{min}}$. With increasing bath temperature these features are washed out. For $T_{\rm{bath}} > $ 50\,K they are not visible anymore indicating that here the concept of a hot spot becomes useless. 
Further note that for $T_{\rm{bath}}$ = 70\,K, the $T_{\rm{min}}$ curve intersects the $T_{\rm{max}}$ curve for $T_{\rm{bath}}$ = 50\,K,  
while the $T_{\rm{max}}$ curve for $T_{\rm{bath}}$ = 70\,K lies almost on top of the $T_{\rm{max}}$ curve for $T_{\rm{bath}}$ = 30\,K. This reflects the fact that temperature differences in the stack are much stronger in the presence of a hot spot than for the more homogeneous case of $T_{\rm{bath}}$ = 70\,K.
The results for the full calculation are shown by symbols.
As for the IVC the symbols are basically on top of the lines for not too low bias currents. For low bias, when the IVC of the stack has switched to an inner branch or to the zero voltage state, $T_{\rm{max}}$ exhibits a jump towards lower temperatures and coincides with $T_{\rm{min}}$ at low bias.  
%
\begin{figure}[tb]
\includegraphics[width=\columnwidth,clip]{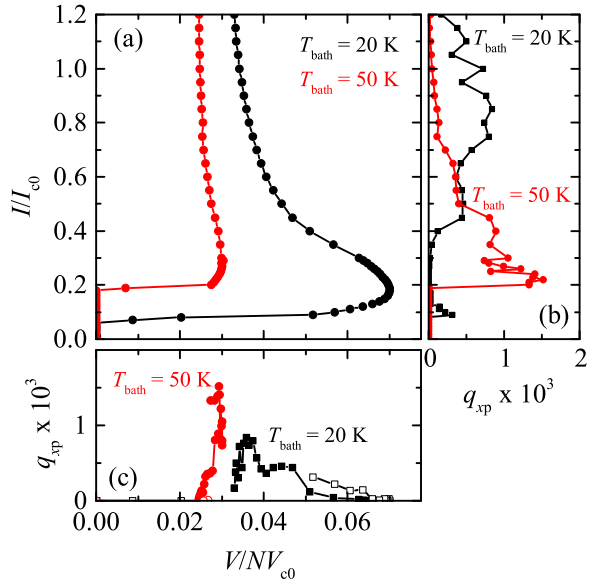}
\caption{(color online) Full calculation results for two selected bath temperatures: (a) IVCs and (b,c) in-plane power dissipation $q_{x{\rm p}}$  vs. bias current (b) and vs. voltage across stack (c).  In (b) and (c) squares are for $T_{\rm{bath}}$ = 20\,K and  circles for $T_{\rm{bath}}$ = 50\,K. In (c) the solid symbols are for the high bias region and open symbols for the low bias region.
}
\label{fig:IVC_Emi}
\end{figure}
%

In Fig.~\ref{fig:IVC_Emi} we show for two selected bath temperatures (20\,K and 50\,K) IVCs, as calculated from the full model together with the in-plane power $q_{x{\rm p}}$. $q_{x{\rm p}}$ was calculated from Fourier transforms of timetraces taken over 512/$v$ time units (about 40 Josephson oscillations), with an elementary step of 0.5/$v$ time units.  
The Fourier spectra were averaged 10 times and integrated from 0.9\,$f_{\rm e}$ to 1.1\,$f_{\rm e}$ where $f_{\rm e}$ is the peak frequency of $q_{x,{\rm av}}(f)$. 
With this ``short'' time Fourier transforms the frequency resolution is low and the linewidth of $q_{x,{\rm av}}(f)$ is in fact much larger than its actual linewidth obtained for long term integration (see below). This resembles the experimental situation where the Fourier spectrometers used for measuring THz emission have a linewidth of several GHz, whereas the real linewidth of the Josephson emission is in the sub-GHz range.  

For $T_{\rm{bath}}$ = 20\,K $q_{x{\rm p}}$ is large mainly at high bias and is peaked at currents near 0.8\,$I_{\rm c0}$, cf. Fig.~\ref{fig:IVC_Emi}(b). 
Note that $q_{x\rm{p}}$ is lower for the values of the normalized bias current of 1.05 and 0.95 than for their adjacent values. This is not due to thermal noise but results from two competing modes in the stack (standing wave patterns) with different wavelengths. 
As a function of voltage the value of $q_{x{\rm p}}$ (at $T_{\rm bath}$ = 20\,K) is large near a normalized voltage of 0.04 per junction, corresponding to a Josephson frequency of about 580\,GHz. At the maximum, $q_{x{\rm p}} \approx 8\cdot10^{-4}$, corresponding to 0.48\,mW.  

For $T_{\rm{bath}}$ = 50\,K, $q_{x{\rm p}}$ rises with decreasing current and has a peak near 0.2\,$I_{\rm c0}$ and for a normalized voltage near 0.03. The peak value is about 0.0015, i.e. by about a factor of 2 larger than the maximum value observed in the 20\,K curve.
%
\begin{figure}[tb]
\includegraphics[width=\columnwidth,clip]{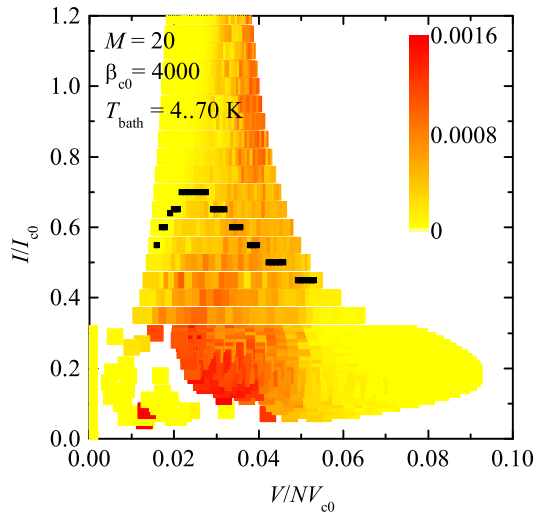}
\caption{(color online) In-plane power $q_{x{\rm p}}$ (color scale) as a function of normalized bias current and normalized voltage across the stack. The bath temperature was varied between 4\,K and 70\,K in steps of 2\,K. Black symbols indicate the currents above (below) which the maximum temperature in the stack was above (below) $T_{\rm c}$ = 85\,K ($T_{\rm c}$ line). 
}
\label{fig:2D_plots_a}
\end{figure}
%
The in-plane power $q_{x{\rm p}}$ shown in Figs.~\ref{fig:IVC_Emi}(b) and (c) exhibits some structure (as it may be similar to the one observed  experimentally for the THz emission power\cite{Wang10a}) but does not give a conclusive picture on how either the power or the structures evolve with bath temperature, dc voltage and bias current. In fact, in experiment one often observes that there is emission both at low bias and at high bias, with the larger power observed in the low bias regime. To obtain a more systematic picture for $q_{x{\rm p}}$, we calculated, for bath temperatures between 4\,K and 70\,K, in steps of 2\,K, a large number of IVCs together with $q_{x{\rm p}}$ for each point and plot it as the color scale in the IVC family, cf. Fig.~\ref{fig:2D_plots_a}.  
In the graph we have also indicated by black symbols the currents above (below) which the maximum temperature in the stack was above (below) $T_{\rm c}$ = 85\,K ($T_{\rm c}$ line).

Fig.~\ref{fig:2D_plots_a} shows that the maximum of $q_{x{\rm p}}$ is in the low-bias regime for currents below 0.5\,$I_{\rm c0}$ and voltages per junction below 0.05 $V_{\rm c0}$. The in-plane power in the high-bias regime is lower and confined in a voltage region between roughly 0.02 and 0.05 $V_{\rm c0}$ per junction. In the low-voltage region at high bias the temperature in the stack is above $T_{\rm c}$  at (almost) each value of $x$. There are almost no supercurrents left in the stack, and the remaining ac Josephson currents do not excite resonant modes. Thus, obviously, $q_{x{\rm p}}$ is low. 
However, at normalized voltages above 0.05 the value of $q_{x{\rm p}}$ is also low, although the temperature in the stack is well below $T_{\rm c}$. This regime will be addressed in more detail in the next subsection. 

For selected bias points we also performed long-term (over 5200 Josephson oscillations) Fourier transforms of $q_{x,{\rm av}}$. It turned out that the linewidth $\Delta f$ of $q_{x,{\rm av}}(f)$ was close to the resolution limit (i.e. the peak in the Fourier transform consisted of only 3 points) both in the high- and the low-bias regime. For example, at $I/I_{\rm c0} = 0.6$ and $T_{\rm{bath}} =$20\,K, $\Delta f <$ 75\,MHz and at $I/I_{\rm c0} = 0.3$ and $T_{\rm{bath}} =$50\,K, $\Delta f <$ 50\,MHz. 
For such small linewidths residual drifts in the temperature distribution in the stack are likely to affect the results and thus we did not go for even longer integration times. We thus cannot make a conclusive statement about $\Delta f$ and its scaling with $T_{\rm{bath}}$ or $M$. However, it seems that the strong broadening of the emission spectra at low bias, which is observed experimentally \cite{Li12}, is not contained in the present model. In the simplified model of Ref. \onlinecite{Gross13} $\Delta f$ became large in the low-bias regime once a spread in the junction critical currents and resistivities was introduced. This is not implemented yet in the present model.  

\subsection{Local properties}
%
\begin{figure}[tb]
\includegraphics[width=\columnwidth,clip]{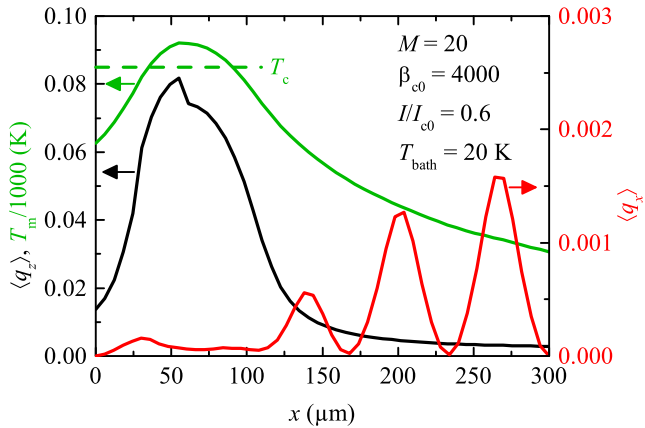}
\caption{(color online) Time averaged distribution of dissipated power density $q_z(x)$ generated by $c$-axis currents (black line, left scale), the temperature $T_{\rm m}(x)$ in the mesa (green line, left scale) and time average of power density of dissipated power $q_x(x)$ generated by in-plane currents (red line, right scale). 
At given $x$ position $q_z$ and $q_x$ have been averaged over all segments.
The bath temperature is 20\,K and the bias current is 0.6\,$I_{\rm c0}$. The noise parameter is $\Gamma_0 = 5 \cdot 10^{-5}$.   
}
\label{fig:i06T20_averages}
\end{figure}
%
We now look, for selected bias points, at local properties. Fig.~\ref{fig:i06T20_averages} shows time averages at $T_{\rm bath}$ = 20\,K and $I$ = 0.6\,$I_{\rm c0}$. 
The temperature distribution $T_{\rm m}(x)$ shows a maximum temperature above $T_{\rm c}$ (a hot spot) located near $x$ = 70\,$\mu$m. This is slightly to the right of the input lead which extends from $x$ = 30\,$\mu$m to $x$ = 60\,$\mu$m. Note that a large fraction of the right hand side of the stack and a small fraction on the left hand side are at temperatures below $T_{\rm c}$. The dissipated power density $\left\langle q_z(x)\right\rangle$ generated by out-of-plane currents, averaged over time and all segments at each position $x$, is also shown in Fig.~\ref{fig:i06T20_averages}. Here we added the heat production of the bias lead which appears as an additional rectangle on top of the Gaussian shaped power density generated by the out-of-plane currents. While $\left\langle q_z(x)\right\rangle$ is smooth, except for the contribution created by the lead, its counterpart for the power density $\left\langle q_x(x)\right\rangle$ generated by in-plane currents exhibits pronouned oscillations, with a wavelength of about 75\,$\mu$m. The oscillations appear for $x > 120$\,$\mu$m, i.e. in the superconducting right hand part of the stack. 

%
\begin{figure}[tb]
\includegraphics[width=\columnwidth,clip]{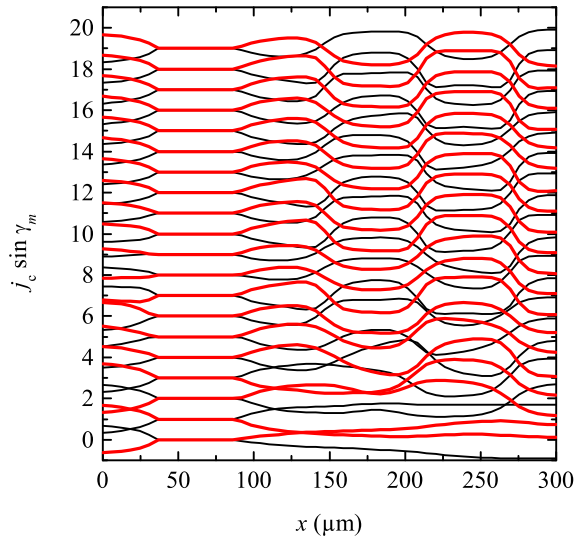}
\caption{(color online) Two snapshots of Josephson currents $j_{\rm c}(T_{\rm m})\sin\gamma_m$ at normalized times $t_1$ (black curves) and $t_2$ (red curves), differing by roughly half of an oscillation period. Snapshots are for $M$ = 20, $I = 0.6 I_{\rm c0}$ and $T_{\rm{bath}} = 20$\,K. Curves for adjacent segments are vertically offset.
}
\label{fig:i06T20_singam}
\end{figure}
%
To obtain more insight into the dynamics of the stack we monitored the time dependence of the in-plane and out-of-plane currents. In the presence of fluctuations these quantities look very noisy. We thus, for demonstration, initialized our simulation with a ``noisy'' set of variables and then turned off the fluctuations, i.e. we set $\Gamma_0$ = 0. Fig.~\ref{fig:i06T20_singam} shows two snapshots of the Josephson currents $j_{\rm c}(T_{\rm m})\sin\gamma_m$. Curves for adjacent segments are vertically offset. Note that in regions with $T_{\rm m} >T_{\rm c}$ the supercurrents are zero. Looking at the  curves at time $t_1$ one notes that at the left superconducting part of the stack the curves of segments 11--20 bend downwards while for segments 1--10 some of the curves are bent upwards and some downwards. On the right hand part of the stack in segments 5--20 there are clear oscillations along $x$, with nodes near $x$ = 148, 209 and 274\,$\mu$m. In fact, the snapshot has been made at a time when the supercurrent to the left of the node at $x$ = 209 \,$\mu$m was near its maximum. The curves at $t_1$  are for a time where the supercurrents at the same position were roughly at their minima. The three nodes of $j_{\rm c}(T_{\rm m})\sin\gamma_m$ in the superconducting part of the mesa and the fact that, left (right) of a given zero, the supercurrents rise to their maximum (minimum), i.e. $\sin\gamma_{\rm m} \rightarrow \pm 1$, indicate that lines of vortices and antivortices have formed. These are chains of $\pi$ kinks\cite{Lin08,Koshelev08b,Hu08, Hu09,Koshelev10,Lin10a,Lin10b,Lin10c} located at the three nodes. It has been conjectured before that such states appear in mesa structures subject to strong self-heating \cite{Kakeya12}. 

%
\begin{figure}[tb]
\includegraphics[width=\columnwidth,clip]{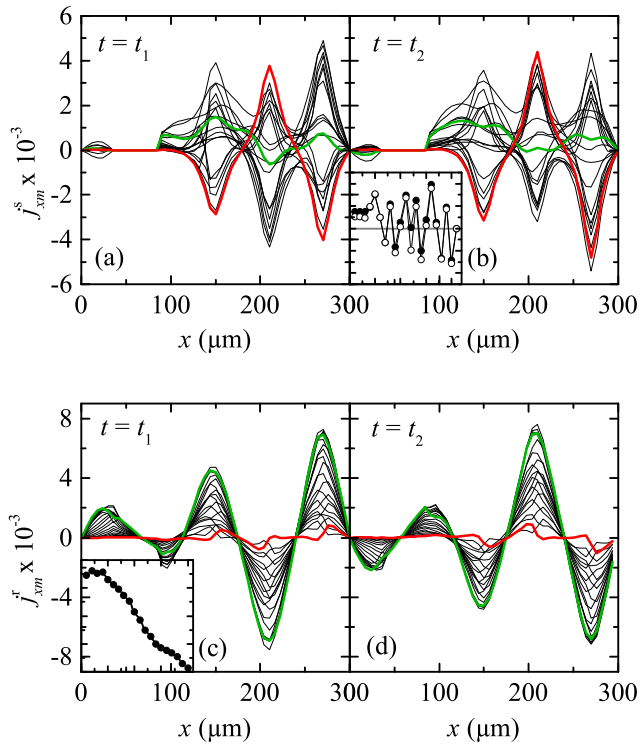}
\caption{(color online) Snapshots of in-plane supercurrent densities $j^{\rm s}_{{x},m}(x)$ at (a) time $t_1$ and (b) time $t_2$. Snapshots of in-plane quasiparticle current densities $j^{\rm r}_{{x},m}(x)$  at (c) time $t_1$ and (d) time $t_2$. Inset in (b) shows values of  $j^{\rm s}_{{x},m}(x)$ at $x$ = 150\,$\mu$m vs. $m$ at times $t_1$ (solid symbols) and $t_1$ (open symbols). Zero value of $j^{\rm r}_{{x},m}$ is indicated by a horizontal line. Inset in (c) shows $j^{\rm r}_{{x},m}(x)$ at $x$ = 270\,$\mu$m vs. $m$ at time $t_1$. In the graphs green (red) lines indicate the current densities in electrodes $m$ = 1 ($m = M$). $I$ = 0.6 $I_{\rm c0}$, $T_{\rm bath}$ = 20\,K and $M$ = 20.
}
\label{fig:i06T20_jx}
\end{figure}
%
Fig.~\ref{fig:i06T20_jx}(a) and (b) show snapshots of the in-plane supercurrent densities $j^{\rm s}_{{x},m}(x)$ at the same times $t_1$ and $t_2$, respectively. The curve for electrode 1 is indicated by a green line, the curve for electrode $M$ is shown by a red line. In contrast to the Josephson current densities of Fig.~\ref{fig:i06T20_singam} the $j^{\rm s}_{{x},m}(x)$ at given $x$ do not reverse sign at $t_1$ compared to $t_2$, i.e. the $j^{\rm s}_{{x},m}(x)$ distribution is almost static, as expected for $\pi$ kink lines. 
In the inset of Fig.~\ref{fig:i06T20_jx}(b) the currents $j^{\rm s}_{{x},m}(x)$ are shown vs. layer index $m$ for $x$ = 150\,$\mu$m and times $t_1$ (solid symbols) and $t_2$ (open symbols). Besides the fact that open and solid symbols are located almost on top of each other one notes that $j^s_{x,m}(x)$ vs. $m$ oscillates between negative and positive values. For several values of $m$, $j^{\rm s}_{{x},m}(x)$ is near zero. In this case a vortex extends over the two segments adjacent to the center layer. 
In contrast to the supercurrents the quasiparticle currents  $j^{\rm r}_{{x},m}(x)$ have basically the same polarity for all layer indices $m$ at given $x$, cf. Fig.~\ref{fig:i06T20_jx}(c) for a snapshot at time $t_1$ and the inset of this figure for a plot of $j^{\rm r}_{{x},m}(x)$ vs. $m$ at $x$ = 270\,$\mu$m. The highest amplitudes of $j^{\rm r}_{{x},m}(x)$ are reached in the uppermost layers (with low values of $m$). For $m > 2$ this amplitude decreases with increasing $m$, i.e. towards the base crystal which we treat as a ground. Fig.~\ref{fig:i06T20_jx}(d) shows a snapshot of $j^{\rm r}_{{x},m}(x)$ at time $t_2$. All curves  $j^{\rm r}_{{x},m}(x)$ have reversed sign compared to Fig.~\ref{fig:i06T20_jx}(c). Note that the oscillations of $j^{\rm r}_{x,m}(x)$ extend across the hot area, i.e. the in-plane resistive currents are coupled across the whole stack. The oscillation in the hot part of the stack is in fact also vaguely visible in $\left\langle q_x(x)\right\rangle \propto \left\langle j^{\rm r 2}_{{x},m}(x) \right\rangle$, cf. Fig.~\ref{fig:i06T20_averages}. 

Figures \ref{fig:i06T20_singam} and \ref{fig:i06T20_jx} demonstrate that, at the bias current and temperature considered, the stack is in a resonant state involving $\pi$ kink states in its dynamics. $j^{\rm r}_{x,m}(x)$ exhibits $k$ = 5 half waves along the stack. The oscillation frequency, cf. Fig.~\ref{fig:i06T20_jx} (a)  is 0.04 $f_{\rm c0}$ = 580\,GHz. This corresponds to a mode velocity $c_1 = 2L_{\rm s}f/k$ = $7 \cdot 10^7$\,m/s -- roughly the value expected from Eq.~(\ref{eq_A:mode_velocity}) if we use an average temperature of 50\,K for the superconducting part of the stack. 

%
\begin{figure}[tb]
\includegraphics[width=\columnwidth,clip]{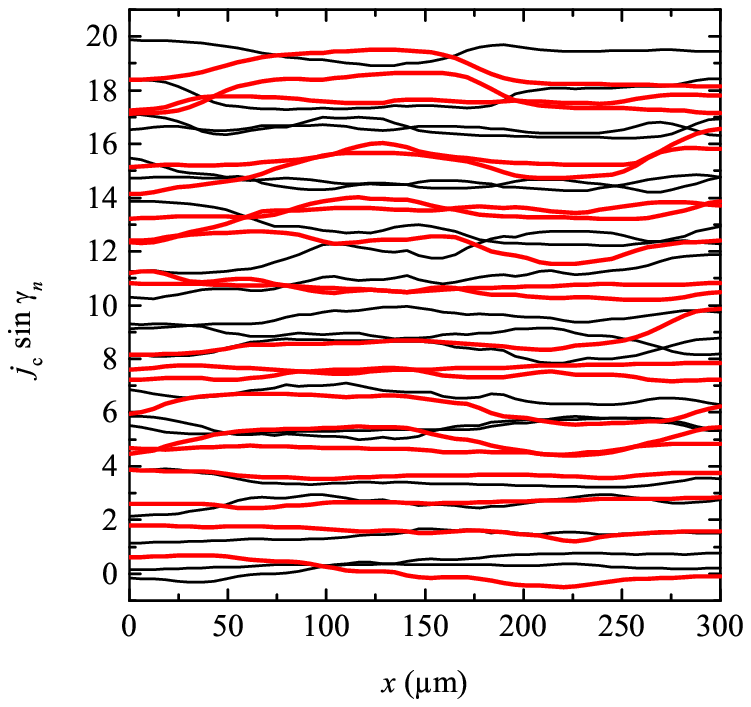}
\caption{(color online) Two snapshots of Josephson currents $j_{\rm c}(T_{\rm m})\sin\gamma_m$ at times $t_1$ (black curves) and $t_2$ (red curves), differing by roughly half of an oscilation period. Data are for $I = 0.3\, I_{\rm c0}$, $T_{\rm{bath}} = 20$\,K and $M$ = 20. Curves for adjacent segments are vertically offset.
}
\label{fig:i03T20_singam}
\end{figure}
%
We next investigate the bias point $I = 0.3$\,$I_{\rm c0}$ at $T_{\rm{bath}}$ = 20\,K. For this bias condition the normalized voltage per junction is $v = 0.06$ and the in-plane power $q_{x\mathrm p}$ is very small, cf. Fig.~\ref{fig:IVC_Emi}. Figure \ref{fig:i03T20_singam} shows snapshots for the Josephson currents at two different times. All curves are very smooth and do not show a sign of synchronization. At least for a homogeneous stack one would associate such a current distribution with a McCumber state. We thus see from Fig.~\ref{fig:i03T20_singam} together with Fig.~\ref{fig:2D_plots_a} that for too high dc voltages across the stack synchronization of the different segments has not been achieved. This basically happens, for the parameters used, for all voltages per junction (normalized frequencies) larger than about 0.05. We conclude that there is a maximum frequency for the resonant modes that can be excited. This frequency in fact depends on the mode velocity. For example, for $G = 35$ and $\beta_{c0} = 4000$ standing waves appeared at least up to $v = 0.7$, as will be shown in section III.C.

%
\begin{figure}[tb]
\includegraphics[width=\columnwidth,clip]{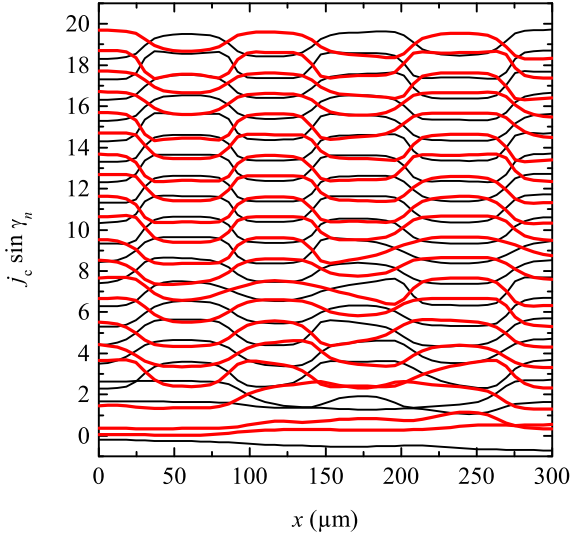}
\caption{(color online) Two snapshots of Josephson currents $j_{\rm c}(T_{\rm m})\sin\gamma_m$ at times $t_1$  and $t_2$, differing by roughly half of an oscillation period. Data are for $I = 0.3$\,$I_{\rm c0}$, $T_{\rm{bath}} = 50$\,K and $M$ = 20. Curves for adjacent segments are vertically offset.
}
\label{fig:i03T50_singam}
\end{figure}
%
\begin{figure}[tb]
\includegraphics[width=\columnwidth,clip]{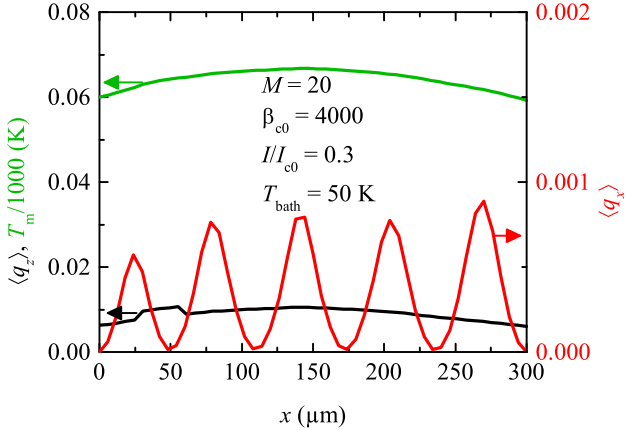}
\caption{ (color online) Time averaged distribution of dissipated power density $q_z (x)$ generated by $c$-axis currents (black line, left scale), the temperature $T_{\rm m} (x)$ in the mesa (green line, left scale) and the time average of power density $q_{x}(x)$ generated by in-plane currents (red line, right scale). 
At given $x$ position $q_z(x)$ and $q_x(x)$ have been averaged over all $M$ segments.
$T_{\rm bath}$ = 50\,K, $I$ = 0.3 $I_{\rm c0}$ and $M$ = 20. The noise parameter is $\Gamma_0 = 5 \cdot 10^{-5}$.   
}
\label{fig:i03T50_averages}
\end{figure}
%
Fig.~\ref{fig:i03T50_singam}  shows two snapshots of Josephson currents $j_{\rm c}(T_{\rm m})\sin\gamma_m$ at normalized times $t_1$ (black curves) and $t_2$ (red curves) for $I = 0.3\,I_{\rm c0}$ and $T_{\rm{bath}} = 50$\,K. Fig.~\ref{fig:i03T50_averages} shows time averages of the power densities $q_z(x)$ and $q_x(x)$ plus the local temperature $T_1(x)$. No hot spot is present at this bias point. The formation of a standing wave is clearly visible in both graphs. In a plot of quasiparticle currents  $j^{\rm r}_{x,m}(x)$  vs. $x$ (not shown) one observes 5 half waves, i.e. essentially the same resonance as in Fig.~\ref{fig:i06T20_jx}. In fact, we have seen similar resonant patterns, with various wave indices $k$,  for many bias currents and temperatures,  whenever a substantial emission was found. The formation of standing waves associated with $\pi$ kink states thus seems to be a robust feature both in the high- and in the low-bias regime. 

%
\begin{figure}[tb]
\includegraphics[width=\columnwidth,clip]{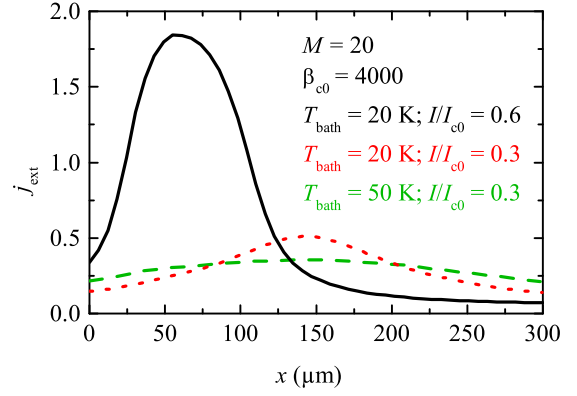}
\caption{(color online) Bias current density $j_{\rm{ext}}(x)$ for  different values of $T_{\rm bath}$ and $I/I_{\rm c0}$.  
}
\label{fig:i03T50_bias}
\end{figure}
%
We finally briefly look at the bias current density $j_{\rm{ext}} (x)$. Fig.~\ref{fig:i03T50_bias}  shows this quantity for the bias conditions discussed above, i.e. 
$T_{\rm{bath}} = 20\,\rm{K}$; $I/I_{\rm c0}$ = 0.6 , $T_{\rm{bath}} = 20\,\rm{K}$; $I/I_{\rm c0}$ = 0.3 and $T_{\rm{bath}} = 50\,\rm{K}$; $I/I_{\rm c0}$ = 0.3. For $T_{\rm{bath}} = 20\,\rm{K}$ and $I/I_{\rm c0}$ = 0.6, i.e. in the presence of a hot spot, most of the bias current flows through the hot region, leading to a very low current density in the ``cold'' parts of the stack. By contrast the current density profiles are much more smooth at the two other bias points and the current density is in fact higher than for the high-bias case. Thus, the bias condition in the cold region is comparable in all cases. For $T_{\rm{bath}} = 20$\,K and $I/I_{\rm c0}$ = 0.6 and $T_{\rm{bath}} = 50$\,K and $I/I_{\rm c0}$ = 0.3 also the dc voltage drop across the stack in the cold part is similar and one may not be very surprized that comparable wave patterns appear. 

\subsection{Scaling behavior}
%
\begin{figure}[tb]
\includegraphics[width=\columnwidth,clip]{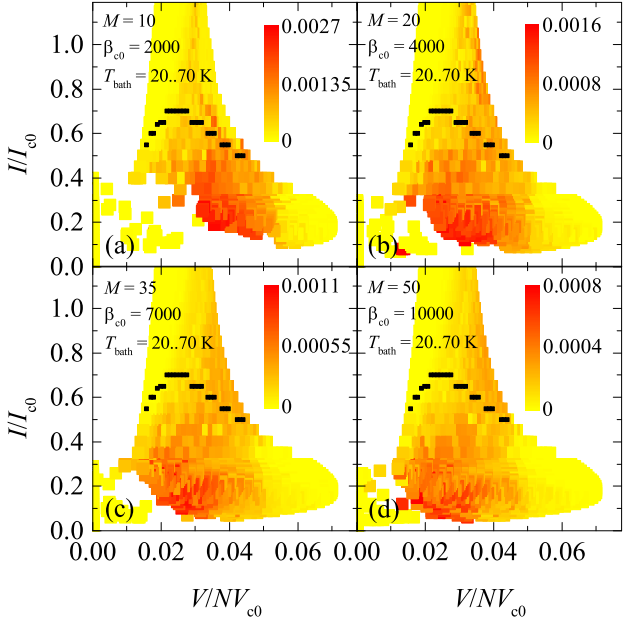}
\caption{(color online) 
In-plane power $q_{x{\rm p}}$ (color scale) as a function of normalized bias current and normalized voltage across the stack for 4 values of $M$ as a function of normalized bias current and normalized voltage across stack. $T_{\rm bath}$ was varied between 20\,K and 70\,K in steps of 2\,K. The product $\beta_{\rm c0}G = 1.4 \cdot 10^5$ is kept constant.
Black symbols denote the $T_{\rm c}$ line. 
}
\label{fig:2D_plots_b}
\end{figure}

%
\begin{figure}[tb]
\includegraphics[width=\columnwidth,clip]{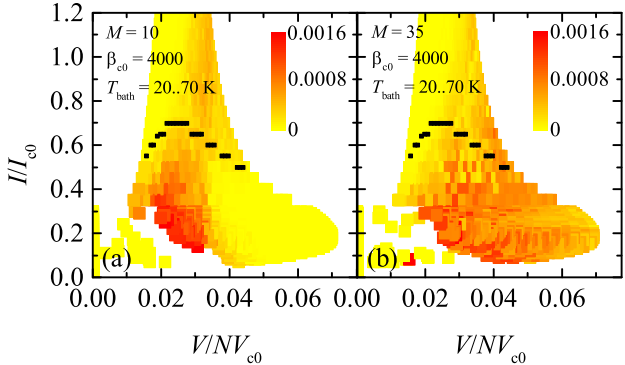}
\caption{(color online) In-plane power $q_{x{\rm p}}$ (color scale) as a function of normalized bias current and normalized voltage across the stack for (a) $M = 10$ and (b) $M = 35$. $\beta_{\rm c0} = 4000 $. $T_{\rm bath}$ was varied between 20\,K and 70\,K in steps of 2\,K. The product $\beta_{\rm c0}G = 1.4 \cdot 10^5$ is kept constant.
Black symbols denote the $T_{\rm c}$ line. 
}
\label{fig:2D_plots_c}
\end{figure}

The results shown above were all for $M = 20$. We next investigate the scaling behavior with $M$. Fig.~\ref{fig:2D_plots_b} shows for $M$ between 10 and 50 the in-plane dissipated power $q_{x{\rm p}}$ (color scale) 
at each point of an IV family taken at bath temperatures between 20\,K and 70\,K in steps of 2\,K.
One notes that plots (b) ($M$ = 20; $G$ = 35; data selected from Fig.~\ref{fig:2D_plots_a}), (c) ($M$ = 35; $G$ = 20) and (d) ($M$ = 50; G = 14) look very similar in the sense that $q_{x{\rm p}}$ is large at about the same voltages and bias currents. For $M = 10$ one observes differences particularly in the high-bias regime where $q_{x{\rm p}}$ is suppressed at high voltages. Here it turned out that no standing wave patterns as discussed in the previous section have formed. Instead, the current densities in the stack fluctuated strongy along $x$. The length scale of these fluctuations (we investigated this also with simulations using up to 200 grid points along $x$) was only a few $\lambda_{\rm k}$. Similar solutions were in fact also found for the $M = 20$ case when the in-plane resistivity $\rho_{ab0}$ was increased by a factor of 5 or more. Thus, for the parameters used in  Fig.~\ref{fig:2D_plots_b} there is a qualitative change between $M = 10$ and $M = 20$. 
Further, between $M = 10$ and $M = 50$ the maximum value of $q_{x{\rm p}}$ 
decreased by a factor of about 3.4.  For $M$ between 20 and 350 we also looked at $q_{x{\rm p}}$ for $I$ = 0.6\,$I_{\rm c0}$ at $T_{\rm bath}$ = 20\,K and for $I$ = 0.3\,$I_{\rm c0}$ at $T_{\rm bath}$ = 50\,K. Here, $q_{x{\rm p}}$ dropped from, respectively, 4.6$\cdot 10^{-4}$ to 2.2$\cdot 10^{-5}$  and from 7.5$\cdot 10^{-4}$ to 2.5$\cdot 10^{-5}$, which is very roughly proportional to $G^{-1}$. For $M$ = 700 ($G$ = 1) one may thus expect maximum values of  $q_{x{\rm p}}$ around 5$\cdot 10^{-5}$, corresponding to 30\,$\mu$W. In all cases the in-plane power dissipation is small compared to the total dc power input, which amounts to, e.g. $\sim 15$\,mW at $I$ = 0.6\,$I_{\rm c0}$, $T_{\rm bath}$ = 20\,K.

For comparison to the data of Fig.~\ref{fig:2D_plots_b}, in Fig.~\ref{fig:2D_plots_c} we show $q_{x{\rm p}}$ as a function of  bias current and  voltage for  $M =10$ and $M = 35$ at a fixed value of $\beta_{\rm c0} = 4000$. In this case the mode velocity $c_1$ is proportional to $M^{1/2}$. As a consequence, for $M = 10$, the region where $q_{x{\rm p}}$ is large, has shifted to lower voltages compared to the case of $M = 20$, and for $M = 35$ it shifted to higher voltages, confirming the proportionality of the regions developing strong standing waves to $c_1$.  

%
\begin{figure}[tb]
\includegraphics[width=\columnwidth,clip]{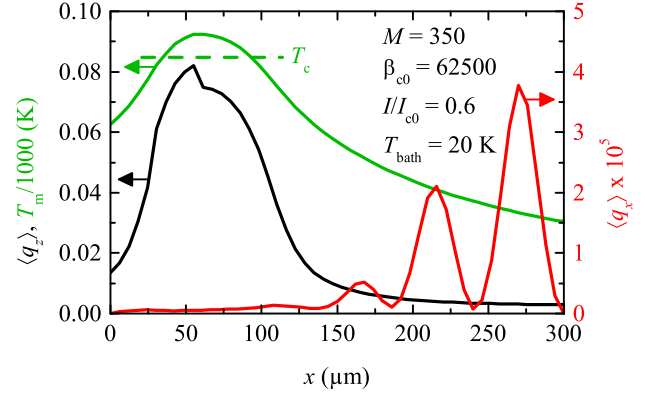}
\caption{(color online) For the case of $M = 350$: time averaged distribution of dissipated power density $q_z(x)$ generated by $c$-axis currents (black line, left scale), the temperature $T_{\rm m}(x)$ in the mesa (green line, left scale) and the time average of dissipated power density $q_x(x)$ generated by in-plane currents (red line, right scale). 
At given $x$ position $q_z$ and $q_x$ have been averaged over all segments.
$T_{\rm bath}$ = 20\,K and the $I$ = 0.6\,$I_{\rm c0}$. The noise parameter is $\Gamma_0 = 5 \cdot 10^{-5}$.   
}
\label{fig:M350_i06T20_averages}
\end{figure}

For $I/I_{\rm c0} = 0.6$ and $T_{\rm{bath}}$ = 20\,K, as well as for $I/I_{\rm c0} = 0.3$ and $T_{\rm{bath}}$ = 50\,K we also calculated current and field distributions for large values of $M$ up to 350 (it was not yet possible to stabilize calculations for $M = 700$). 
The time averaged power densities and the temperature profile were very similar to the ones shown in Figs.~\ref{fig:i06T20_averages} and \ref{fig:i03T50_averages}. Fig.~\ref{fig:M350_i06T20_averages} shows an example for $M = 350$ and $I/I_{\rm c0} = 0.6$,  $T_{\rm{bath}}$ = 20\,K. The graph looks almost identical to Fig.~\ref{fig:i06T20_averages}, with the main difference that $\left\langle q_x\right\rangle$ has decreased by a factor of about 40.

\subsection{Comparisons to experiment}
We now compare our theoretical results to experimental findings, with respect to the role of the hot spot position and LTSLM imaging of standing wave patterns. We will also comment on THz emission properties in relation to the dissipated in-plane power.

In our simulations we have modelled the heat produced by the bond wire by an additional heat source located near the left edge of the stack. In experiment the hot spot has been moved along $x$ by using two current injection leads on the mesa and biasing them with different ratios of currents \cite{Guenon10}. This method has also been used in simulations \cite{Gross12}. It further has been shown in experiment that the appearance of standing wave patterns and the emission power depend on the position of the hot spot \cite{Guenon10,Minami14}. In the present model we performed some calculations at $M = 20$ and $I/I_{\rm c0} = 0.6$,  $T_{\rm{bath}}$ = 20\,K where we varied the position of the hot spot simply by changing the location of the current lead. Standing waves appeared for basically all hot spot positions and the average voltage was almost independent of its position. However, $q_{x{\rm p}}$ was lower by about a factor of 2 when the hot spot was located at the center of the stack. Further, for several positions of the bond wire we observed the competition of two different wave patterns, leading to fluctuations in $q_{x{\rm p}}$ as a function of the position of the bond wire.

In the simulations standing waves were a robust feature over a wide range of bias currents and bath temperatures. By contrast, in LTSLM imaging \cite{Wang09a,Guenon10,Wang10a} standing wave patterns are seen much less often, indicating that the simulations overestimate the stability of cavity resonances. We also attempted to model LTSLM images by introducing an additional heater representing the LTSLM laser beam and monitoring the beam-induced changes $\Delta V(x_{\rm b})$ of the dc voltage across the stack vs. laser beam position $x_{\rm b}$. We did obtain a response proportional to the beam-induced changes of the $c$-axis conductance as simulated previously\cite{Gross12}, however there were only very faint signatures of the wave patterns. This is likely to be caused by the fact that we inject the bias current $j_{\rm{ext}}$ purely along the $c$-axis proportional to the local $c$-axis conductance. 
In more detail, one may obtain $\Delta V(x_{\rm b})$ from a power balance equating the change in input power $I\Delta V$ and the sum of the (time averaged) changes in the integrated in-plane and out-of-plane dissipation
\begin{equation}
\label{eq:LTSLM}
\begin{split}
I \Delta V = \Delta \left\langle\int d(x,y,z) (q_{z} + q_{x})\right\rangle \\
= \int d(x,y,z) (\left\langle \Delta q_{z} \right\rangle + \left\langle \Delta q_{x}\right\rangle)~.
\end{split}
\end{equation}
The brackets denote time averaging and the integration is over the stack volume. As discussed above, $\left\langle q_x(x)\right\rangle$ exhibits clear modulations but is small compared to $\left\langle q_z(x)\right\rangle$. Snapshots of $\left\langle q_z(x)\right\rangle$ indeed exhibit time-dependent wavy modulations (not shown). However, they occur on a large static background representing the dc power in the stack and are thus averaged out in $\left\langle q_z\right\rangle$. To obtain a significant wave signature in LTSLM either $\left\langle q_z\right\rangle$ should exhibit strong modulations or some in-plane voltage should be picked up in the measurements. The latter is the likely scenario and consistent with previous observations that the observed patterns are associated with magnetic fields rather than electric fields \cite{Guenon10}. 

\begin{figure}[tb]
\includegraphics[width=\columnwidth,clip]{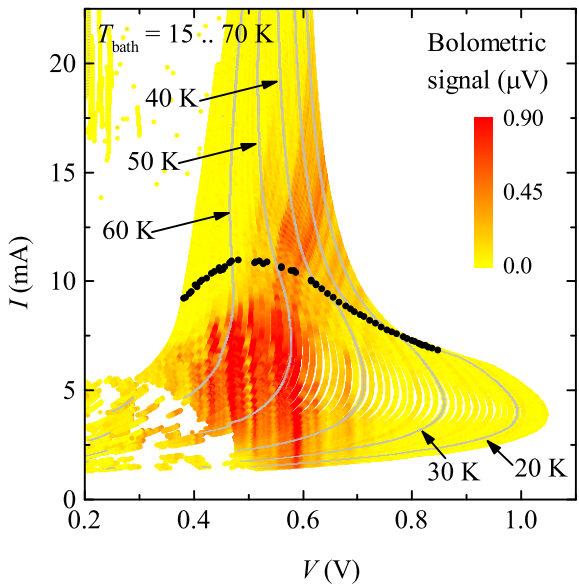}
\caption{(color online) Experimental data for a 165 $\times$ 60 $\mu \mathrm{m}^2$ large z-type IJJ stack consisting of $N = 480$ IJJs: Emitted power (color scale) as a function of bias current and voltage across the stack. The bath temperature was varied between 15\,K and 70\,K in steps of 0.5\,K. Black symbols indicate the currents where a hot spot has formed. The contact resistance is subtracted from each IVC.
}
\label{fig:2dmeasurement}
\end{figure}

We finally comment on the relation of the THz emission power $P_e$ observed in experiment and the absorbed in-plane power $q_{x\mathrm p}$ calculated numerically. In general, these quantities can be very different. For example, if an out-of-phase mode has formed across the stack, $q_{x\mathrm p}$ can be large while $P_e$ will be very small~\cite{Krasnov10}. On the other hand, for in-phase modes, as we see them in our simulations, a major part of the in-plane currents contributes constructively to the ac magnetic field produced at the boundaries of the stack. In fact, preliminary simulations performed with radiative boundary conditions indicate that the emitted power is on the order of 5--10\% of the absorbed ac power. In the absence of a standing wave, like for the McCumber-like state shown in Fig.~\ref{fig:i03T20_singam}, both the ac electric field and the ac magnetic field at the boundaries are small and we expect both $q_{x\mathrm p}$ and the emitted power to be small as well. 
With these caveats in mind, let us look at some experimental data for $P_e$. The experimental findings will be discussed in detail in a separate publication. The measurements have been performed on a 165 $\times$ 60 $\mu\mathrm{m}^2$ large z-type IJJ stack~\cite{Yuan12} consisting of $N = 480$ IJJs. The electrothermal behavior of such z-type stacks is in fact similar to mesa structures. Fig.~\ref{fig:2dmeasurement} shows $P_e$ as the color scale for a large family of IVCs taken at bath temperatures between 15\,K and 70\,K. On a qualitative level this plot can be compared to Fig.~\ref{fig:2D_plots_a} or to Fig.~\ref{fig:2D_plots_b}.

In dimensioned units the maximum voltage per junction in Fig.~\ref{fig:2D_plots_b} at 20\,K is about 2.1\,mV and is in line with the corresponding 20\,K value of 2.1\,mV in Fig.~\ref{fig:2dmeasurement}. In the experimental data the highest emission signals are seen in the absence of a hot spot for voltages between 0.6\,V and 0.4\,V, corresponding to, respectively, 1.25\,mV and 0.8\,mV per IJJ. On a somewhat lower power level the emission continues into the hot spot region. For voltages larger than 0.7\,V the emission is basically absent.

Comparing the color distributions in Fig.~\ref{fig:2D_plots_b} and Fig.~\ref{fig:2dmeasurement} the only qualitative, but striking, difference is the appearance of a stripe-like modulation of $P_e$ in Fig.~\ref{fig:2dmeasurement}. In the low bias regime these stripes appear at about constant voltage while in the high-bias regime they are tilted to the right. A detailed discussion of the stripes -- we find similar patterns also for other structures like bare IJJ stacks embedded between Au layers -- is out of the scope of this paper. We just briefly mention here that, by measuring the frequency of emission $f_e$ using a superconducting integrated receiver~\cite{Li12}, we found that also in the high-bias regime $f_e$ is almost constant along a given stripe. The reason for the observed tilt of the stripes is that in the presence of a hot spot there is an additional dc voltage due to in-plane currents, which is picked up by the voltage leads.

Further, the stripe features are most likely not due to different resonant modes inside the IJJ stack but are probably caused by extrinsic effects like interferences within the substrate or other parts of the setup.

\section{Summary}
\label{Sec:Conclusions}
In summary, by using 1D coupled sine-Gordon equations combined with heat diffusion equations, we have numerically investigated the thermal and electromagnetic properties of a 300\,$\mu$m long intrinsic Josephson junction stack consisting of $N$ = 700 junctions. The junctions in the stack were combined to $M$ segments. We assumed that inside each segment all junctions behave identically. Most simulations were performed for $M = 20$, i.e. each segment consisted of $N/M$ = 35 junctions. The thermal properties were (by ansatz and the fact that $\left\langle q_x\right\rangle$ $\ll$ $\left\langle q_z\right\rangle$) basically independent of $M$ and showed the appearence of a hot spot for high bias currents. For $M$ between 20 and 50 local current and electric field distributions were similar, provided that the mode velocity in the stack was chosen to be independent of $M$. In particular, robust standing wave patterns appeared and were identified to be associated with $\pi$ kink states. For two bias points we confirmed that the same waves/$\pi$ kink states are also present for $M = 350$. For the case of $M = 10$ (and also lower values of $M$) solutions with strongly fluctuating currrent densities and electric fields appeared at high bias currents, replacing the long-wavelength standing wave patterns.  This causes some problems with models treating intrinsic Josephson junction stacks as a single ``giant'' junction ($M = 1$), since the formation of $\pi$ kinks requires at least 2 segments. For modest values of $M$ (e.g. $M = 2$) there may be some effective values of the parameters to reproduce the physics of the large IJJ stacks. However, at least for the parameter values and boundary conditions used in this work, $M$ should be 20 or larger.
 
Also, one should be cautious when extrapolating the results to $M = N$ and to a real 3D situation. There might be  qualitative changes in electromagnetic behavior, as they have been seen in our simulations for high bias currents and segment numbers between $M$ = 10 and $M$ = 20. With respect to 3D stacks it has been found in experiment that emission properties scale inversely proportional with the width of the stack, indicating that also a standing wave has formed along the stack width. This can not be described within the 1D coupled sine-Gordon equations. An extension to the 2D version of these equations remains to be done.

%

%
\acknowledgments
We gratefully acknowledge financial support by the National Natural Science Foundation of China (Grant No. 11234006), the Priority Academic Program Development of Jiangsu Higher Education Institutions, the Deutsche Forschungsgemeinschaft (Project KL930/12-1), the Grants-in-Aid for scientific research from JSPS, and RFBR grants 13-02-00493-a and 14-02-91335 and Ministry of Education and Science of the Russian Federation (No. 14.607.21.0100). 

\appendix* 

\section{Basic equations and numerical details}
 \label{Sec:Appendix} 
 
\subsection{Basic equations}

The general geometry of the mesa structure under consideration has been introduced in Sec. \ref{sec:model}A, cf.~Fig.~\ref{fig:geometry}. \\ 
\\ \textbf{Thermal description}

As stated in Sec. \ref{sec:model}A we write for the heat diffusion in the $k$th layer:
\begin{equation}
\label{eq_A:heat_nth_layer}
c_k\dot{T}_k=\frac{d}{dx}\left(\kappa_{\|,k}\frac{d}{dx}T_k\right)+\frac{2}{D_k}\left(j_{{\rm{in}},k}-j_{{\rm{out}},k}\right)+q_k,
\end{equation}
We assume that $T_k$ is the temperature in the center of layer $k$ along $x$. Introducing the auxiliary temperature $T_{{\rm h},k}$ as the temperature at the interface between layers $k$ and $k-1$ one obtains for the heat current densities
\begin{equation}
\label{eq_A:heatcurrent_in}
j_{{\rm{in}},k} = \kappa_{\bot,k}(T_{{\rm h},k}-T_k)\frac{2}{D_{k}}
\end{equation}
and
\begin{equation}
\label{eq_A:heatcurrent_out}
j_{{\rm{out}},k} = \kappa_{\bot,k}(T_{k}-T_{{\rm h},k+1})\frac{2}{D_{k}},
\end{equation} 
where $\kappa_{\bot,k}$ is the out-of-plane thermal conductivity of layer $k$. We demand $j_{{\rm{in}},k} =j_{{\rm{out}},k-1}$, yielding 
$ T_{{\rm h},k} =(a_kT_k+a_{k-1}T_{k-1})/(a_k+a_{k-1})$, with $a_k=\kappa_{\bot,k}/D_k$.

For the mesa including the Au layer ($k = 0$) $T_0 = T_{\rm m}$. There is no heat flow into this layer and thus $T_{{\rm h},0}=T_{\rm m}$ here. 
For the effective thermal conductivity of the mesa/gold layer, since for in-plane heat flow the BSCCO stack and the Au layer are in parallel, we take the weighted average
\begin{equation}
\label{eq_A:kappa_eff}
\kappa_{\|,0} = \frac{D_{\rm m}\kappa_{ab}+D_{\rm{Au}}\kappa_{\rm{Au}}}{D_{\rm m}+D_{\rm{Au}}}
\end{equation} 
of the BSCCO in-plane thermal conductivity $\kappa_{ab}$ and the gold thermal conductivity $\kappa_{\rm{Au}}$. The perpendicular heat conductivity is limited by the BSCCO $c$-axis thermal conductivity $\kappa_{c}$ and we thus use $\kappa_{\bot,0} = \kappa_{c}$ for this layer. During simulations it has turned out that the effect of self-heating is too strong compared to full 3D simulations as described in Ref. \onlinecite{Gross12}. The reason is that in our 1D scenario there is no heat flow in a given layer along the $y$ direction (we have confirmed this by 3D simulations of a very narrow structure). To compensate for this we decreased the effective mesa thickness by a factor 2 compared to its real thickness in  Eq.~(\ref{eq_A:heat_nth_layer}). 
An expression for the Joule heat power density $q_{\rm m}$ produced in the mesa will be given below, cf. Eq.~(\ref{eq_A:Joule_heat}). 
For the total heat generation we add a contribution by the contacting bond wire, expressed as 
\begin{equation}
\label{eq_A:Joule_heat_wire}
q_{\rm B} = \left\langle j_{\rm{B}}\right\rangle^2\rho_{\rm B} f(x),
\end{equation}
where $\rho_{\rm B}$ is some effective resistivity associated with the wire, having a diameter $L_{\rm B}$ along $x$. $ \left\langle j_{\rm{B}}\right\rangle = I/L_{\rm B}$ is the spatially averaged applied current density.  The function $f(x)$ equals 1 in the wire and is zero outside.  

In the layers representing the base crystal and the glue there is no heat generation, i.e. $q_k = 0$ for all $k \neq$ 0. These layers have a length $L_{\rm b}$ which we have taken as 2$L_{\rm s}$. The mesa is centered above the base crystal. For the layer interfacing the mesa ($k$ = 1) the heat flow current density $j_{\rm{in},1}$ is nonzero only above the mesa. For the bottom of the glue layer ($k$ = $K$+1) we have the boundary condition $T_{{\rm h},K+2} = T_{\rm bath}$ and, finally, for the in-plane heat flow, von Neumann boundary conditions are used, i.e. we assume that no heat is transported through the boundaries of the mesa and the base along $x$.

For further calculations we normalized time to $\Phi_0/2\pi j_{\rm c0}\rho_{c0}s$. Power densities are normalized to $j_{\rm c0}^2\rho_{c0}$, electric fields to $j_{\rm c0}\rho_{c0}$, current densities to  $j_{\rm c0}$ and resistivities to $\rho_{c0}$. This leads to heat capacities normalized to $\Phi_0j_{\rm c0}/2\pi s$ (in units of K$^{-1}$) and heat conductivities normalized to $j_{\rm c0}^2\rho_{c0}$ (in units of $\mu$m$^2$/K).\\ 
\\\textbf{Electrical circuit}

The electric circuit is sketched in Figs.~\ref{fig:geometry}(c) and (d). 
Let us consider a piece of the \textit{n}th IJJ, located between $x$ and $x$ + d$x$.  
For the current flow along $z$, we use an RCSJ type description, i.e. we consider a Josephson current with critical current $I_{{\rm c},n} = j_{{\rm c},n}$d$xW$, a resistive component with $R_{{c},n} = \rho_{{c},n}s/$d$xW$ and a capacitive component with $C_n=\epsilon_n\epsilon_0$d$xW/s$. Nyquist noise is considered via  a current source producing a random current $I^{\rm N}_{{z},n}= j^{\rm N}_{{z},n}$d$xW$ with spectral power density $4k_{\rm B}T_{\rm m}/R_{{c},n}$.
For simplicity, we will assume that $\rho_{{c},n}$, $j_{{\rm c},n}$ and $\epsilon_n$ are the same for all junctions, i.e. we omit the index $n$. 
The in-plane current flow in the $n$th BSCCO layer is described by a resistive component $R_{{ab},n} = \rho_{{ab},n}$d$x/Wd_{\rm s}$ and an inductive component $L_{{ab},n}=\mu_0\lambda_{ab}^2$d$x/Wd_{\rm s}$.  $L_{{ab},n}$ is the kinetic inductance associated with in-plane supercurrents. Here,  $\lambda_{ab}$ is the in-plane magnetic penetration depth of BSCCO. We also consider an in-plane noise current $I^{\rm N}_{{x},n} = j^{\rm N}_{{x},n}d_{\rm s}W$  with spectral power distribution  $4k_{\rm B}T_{\rm m}/R_{{ab},n}$. 
Note that we have neglected the geometric inductance $L_{\rm g} \approx \mu_0s$d$x/W$ in the superconducting in-plane current paths. There are several reasons for this. First, this inductance, for the parameters we are interested in, is much smaller than the kinetic inductance. Second, in the standard description of IJJ stacks it leads to an out-of-plane length scale $\lambda_{c} = [\Phi_0/(2\pi\mu_0 j_{\rm c0}s )]^{1/2} \approx 300$\,$\mu$m which is comparable to the length of the mesa studied here ($300$\,$\mu$m). Thus, even if the kinetic contributions cancel, the finite value of $\lambda_{c}$   would not lead to long-junction effects. It is thus safe to ignore the geometric inductance. Further, from a more practical point of view, if $L_{{\rm g},n}$  were included in the superconducting path it should also be considered for the resistive in-plane paths which would strongly complicate the resulting sine-Gordon like equations.

Using Kirchhoff's laws, after some math one finds for the $n$th IJJ: 
\begin{equation}
\label{eq_A:sigo_single}
\begin{split}
sd_{\rm s}\left(\frac{\dot{\gamma}^\prime_n}{\rho_{ab}}\right)^\prime +d_{\rm s}\left(j^{\rm N}_{{x},n+1}-j^{\rm N}_{{x},n}\right)^\prime + \lambda_{\rm k}^2 \left(n_{\rm s}\gamma^\prime_n\right)^\prime = \\
2j_{z,n}-j_{z,n+1}-j_{z,n-1}.
\end{split}
\end{equation}
Here, the index $n$ runs from 1 to $N$. The various parameters and normalizations have been already introduced in Sec. \ref{sec:model}A.

For the current densities $j_{z,n}$ in $z$ direction one finds
\begin{equation}
\label{eq_A:RCSJ}
j_{z,n} = \beta_{\rm c0} \ddot{\gamma}_n + \frac{\dot{\gamma}_n}{\rho_{{c},n}} + j_{\rm c} \sin(\gamma_n) +j^{\rm N}_{{z},n},
\end{equation}
with $\beta_{\rm c0} = 2\pi j_{\rm c0}\rho_{c0}^2\epsilon\epsilon_0s/\Phi_0$. 

In Eq.~(\ref{eq_A:sigo_single}) $\rho_{ab}$  and $n_{\rm s}$ depend on temperature and thus, in general on $x$. For spatially constant parameters one would obtain a term $\propto \dot{\gamma}^{\prime\prime}_n/\rho_{ab}$  (dissipation due to in-plane currents) and a term $\propto \gamma^{\prime\prime}_n$ which are familiar from the coupled sine-Gordon equations \cite{Sakai93,Kleiner94b}. The term describing in-plane dissipation is often neglected. However, as it has been pointed out in Ref. \onlinecite{Lin12} it plays a crucial role for the synchronization of large IJJ stacks. Further note that, at $T_{\rm c}$, $n_{\rm s}$ as well as $j_{\rm c}$ go to zero. Then, Eq.~(\ref{eq_A:sigo_single}) has no contributions arising from supercurrents anymore. In Refs. \onlinecite{Asai13,Asai14} the temperature dependence of $n_{\rm s}$ has been missed.  

The calculations presented below are for a huge junction number $N = 700$. For  such $N$ it is basically hopeless to solve Eq.~(\ref{eq_A:sigo_single}) on a reasonable time scale.  Let us thus, before proceeding with expressions for the in-plane currents and for the normalized noise currents, introduce the concept of segments consisting of $G$ IJJs. 
We divide the $N$ junction stack into $M$ segments, each consisting of $G = N/M$ junctions (note that $N$ must contain $M$ as a factor to make $G$ integer). Being interested in dynamic states where all junctions oscillate coherently we assume that, within a segment, the $\gamma_n$ and their derivatives are the same.   
Summing up over all junctions Eq.~(\ref{eq_A:sigo_single}) turns into 
\begin{equation}
\label{eq_A:sigo_segment}
\begin{split}
Gsd_{\rm s}\left(\frac{\dot{\gamma}^\prime_m}{\rho_{ab}}\right)^\prime +d_{\rm s}\left(j^{\rm N}_{{x},m+1}-j^{\rm N}_{{x},m}\right)^\prime + G\lambda_{\rm k}^2 \left(n_{\rm s}\gamma^\prime_m\right)^\prime = \\
2j_{{z},m}-j_{{z},m+1}-j_{{z},m-1}.
\end{split}
\end{equation}
The index $m$ runs from 1 to $M$. Eq.~(\ref{eq_A:sigo_segment}) has almost the same form as Eq.~(\ref{eq_A:sigo_single}), but with an effective length $\lambda_{\rm k} \propto G^{1/2}$ and an effective $\rho_{ab} \propto G^{-1}$. 

\begin{figure}[tb]
\includegraphics[width=\columnwidth,clip]{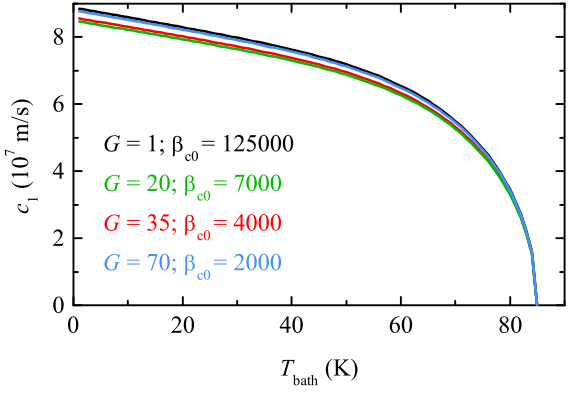}
\caption{(color online) Calculated mode velocity $c_1$ vs. bath temperature in the limit of vanishing dissipation for $G = 1$  $(M = N = 700)$, $G = 20$ $(M = 50)$, $G = 35$ $(M = 20)$ and $G = 70$ $(M = 10)$. For $G = 1$, $\beta_{\rm c0} = 1.25\cdot10^5$, for the other curves $\beta_{\rm c0}G = 1.4\cdot10^5$. }
\label{fig_A:mode_velocities}
\end{figure}
%

For Eq.~(\ref{eq_A:RCSJ}) no modification seems to be necessary, except for the fact that the index $n$ should be replaced by $m$ labelling the segments. There is, however, a subtle point. As shown in the main text, resonant modes appeared in the solutions of Eq.~(\ref{eq_A:sigo_segment}).
For a \textit{homogeneous} $M$ junction stack in the limit $\rho_{ab} \rightarrow \infty$ one can find a set of $M$ collective resonances \cite{Kleiner94,Sakai94}, associated with mode velocities $c_m$, with $m$ = 1...$M$. The resonance where all junctions oscillate in-phase has $m$ = 1 and, rewritten for the segmentation and the boundary conditions we use, has the form\cite{Benseman11} 
\begin{equation}
\label{eq_A:mode_velocity}
c_1 = 2\pi f_{\rm pl} \lambda_{\rm J} \sqrt{G}\frac{1}{\sqrt{1-2\tilde{s}\cos\frac{\pi}{2M+1}}},
\end{equation}
where $\lambda_{\rm J} \approx \lambda_{\rm k}/\sqrt2$ for $\lambda_{c} \gg \lambda_{\rm k}$ and $\tilde{s} = [2+Gd_{\rm s}sn_{\rm s}/\lambda_{ab}^2]^{-1}$. 
The small term $Gd_{\rm s}sn_{\rm s}/\lambda_{ab}^2$ originates from the geometric inductance of the superconducting layers which we neglect in our simulations. We thus work in the limit $s = 0.5$. $f_{\rm pl}$ is the Josephson plasma frequency. 
In our notation  frequencies are normalized to $f_{\rm c0} = \Phi_0/(j_{\rm c0}\rho_{c0}s)$. Then, with $(f_{\rm c0}/f_{\rm pl})^2 = \beta_{\rm c}$, under the condition $N \gg M \gg 1$ we have (at $T$ = 4.2\,K)
$c_1 \approx 4M\lambda_{\rm k}\sqrt{G}f_{\rm c0}/\sqrt{\beta_{\rm c0}} = 4N\lambda_kf_{\rm c0}/\sqrt{\beta_{\rm c0}G}$. Thus, to keep physics independent of $M$,
$c_1$ should be kept constant while changing $M$ and thus (approximately) the product $\beta_{\rm c0}G$ should be kept constant. 

This procedure, however, leads to too strong ac electric fields and particularly the in-plane resistive currents get too strong. In Eq.~(\ref{eq_A:sigo_segment}) we have assumed that $\dot{\gamma}^\prime_n$ is the same for all $G$ junctions inside a segment. The term $\propto \dot{\gamma}^\prime_n$  originates from the difference of the resistive in-plane currents of layers $n$ and $n+1$ and thus the in-plane currents inside a segment should be extrapolated linearly between in-plane currents of the outermost layers in the segment. This results in an extremely high in-plane power dissipation. As a consequence, for example the temperature in the stack shows a strong spatial modulation, of order of some 10\,K, imprinted by the standing wave.
This is clearly not seen in experiment. Further, in simulations we found that the in-plane power density is strongly suppressed when increasing $M$ (and thus $\beta_{\rm c0}$). Thus, in order to return to physical properties which scale reasonably with $M$ we reduced the in-plane power dissipation by considering only the contributions of the outermost layers in a segment. Example simulations for very large numbers of $M$ up to 350 showed that the wave patterns and other properties scale reasonably with this prodedure.
Fig.~\ref{fig_A:mode_velocities} shows $c_1$ vs. $T_{\rm{bath}}$ for $G = 1$ $(M = N = 700)$, $G = 20$ $(M = 50)$, $G = 35$ $(M = 20)$ and $G = 70$ $(M = 10)$. For $G = 1$ $\beta_{\rm c0} = 1.25\cdot10^5$, for the other curves we have used $\beta_{\rm c0}G = 1.4\cdot10^5$.

Thus, for the power dissipation $q_{\rm m}$ we use (in dimensioned units) the expression 
\begin{equation}
\label{eq_A:Joule_heat}
q_{\rm m} = \left(Gs\sum_{m=1}^M \frac{E^2_{z,m}}{\rho_{c}} +d_{\rm s}\rho_{ab}\sum_{m=1}^{M+1}j^{\rm r 2}_{{x},m}\right)\frac{1}{Ns} 
\end{equation}
The first term on the right hand side represents Joule heat generation in the BSCCO stack due to out-of-plane currents, with the electric field $E_{z,m}$ across one of the IJJs in segment $m$. In normalized units, $E_{z,m}$ is replaced by $\dot{\gamma}_m$.  
The second term represents the in-plane dissipation, with the resistive current densities $j^r_{{x},m}$ flowing in the $m$th superconducting layer. The generated power density is averaged over the mesa thickness $Ns$.

We turn to explicit expressions for the in-plane currents. Using, for a piece of length d$x$ of the $m$th superconducting layer, London's equation $E_{{x},m} = \mu_0 \lambda_{ab}^2\dot{j}^s_{{x},n}$, where $E_{{x},m}$ is the in-plane electric field and relating the in-plane voltage drop $E_{{x},m}$d$x$ to the time derivative of the phase $\phi_m$ of the superconducting wave function in this electrode,  $E_{{x},m}$d$x$ = $\Phi_0(\dot{\phi}_m(x+$d$x)-\dot{\phi}_m(x))/2\pi$, we find in our normalized units 
\begin{equation}
\label{eq_A:jxs}
j^{\rm s}_{{x},m} = \frac{\lambda_{\rm k}^2}{d_{\rm s}}n_{\rm s}\phi_m^\prime
\end{equation}
and for the resistive currents experiencing the same electric field one obtains
\begin{equation}
\label{eq_A:jxr}
j^{\rm r}_{{x},m} = \frac{e_{{x},m}}{\rho_{ab}} = \frac{s}{\rho_{ab} } \dot{\phi}_m^\prime
\end{equation}
The index $m$ runs from 1 to $M$+1 and refers to the  CuO$_2$ layers terminating a segment. $e_{{x},m}$ is the normalized in-plane electric field. The $\phi_m$ and $\gamma_m$ are related via
\begin{equation}
\label{eq_A:phi_gamma}
\gamma_m = \frac{\phi^\prime_{m+1}-\phi^\prime_{m}}{G}, 
\end{equation}
allowing to evaluate the in-plane currents once all $\gamma_m$ and in addition $\phi_m^\prime$ of one of the outermost electrodes are known. Note the above relations hold both for the individual layers in the $N$ junction stack (using $n$ instead of $m$) as for the electrodes addressed in the segmented stack.  

This brings us to the boundary conditions. We treat the base crystal as a ground, i.e. we demand that for segment $M$ in Eq.~(\ref{eq_A:sigo_segment}) $j_{{z},m+1} =j_{{z},m}$. Then, the in-plane currents in the superconducting layer $M+1$ are zero, leading to $\phi_{M+1}^\prime = 0$. 
The boundary condition for the $c$-axis currents in junction 1 interfacing the Au layer, i.e. the determination of the current density $j_{{z},m-1}$  in Eq.~(\ref{eq_A:sigo_segment}), is more problematic. In a first attempt  we have explicitly also considered the Au layer as an additional resistive element in parallel to the in-plane resistor describing the quasiparticle currents in the topmost BSCCO CuO$_2$ layer. For realistic resistivities this led to heavy numerical instabilities. We next considered the Au layer to be an ideal conductor. The approach worked, however converged with reasonable computing times only if a resistive layer -- i.e. a contact resistance -- for the out-of-plane currents was introduced between the topmost BSCCO electrode and the Au layer. We thus decided for a simplified description where we assume that the bias current can freely disperse in the Au layer and perhaps some incoherent layer on top of the BSCCO mesa 
and finally enters the topmost IJJ according to the local $c$-axis conductivity, i.e. we used  
\begin{equation}
\label{eq_A:j_ext}
j_{\rm{ext}} = \frac{\left\langle j_{\rm{ext}}\right\rangle}{\left\langle \sigma_{c}\right\rangle\rho_{c}},
\end{equation}
where the brackets denote spatial averaging and $\sigma_{c} = \rho_{c}^{-1}$. 

For the inplane-boundary conditions we used $\gamma^\prime_m (0) = \gamma^\prime_m (L_{\rm s}) = 0$, i.e. there is no energy flow out of the stack along $x$. 


%
%
\bibliography{hot-spots-waves_v2}
\end{document}